\documentclass[twocolumn,superscriptaddress,amsmath,amssymb,aps,prc]{revtex4-1}

\usepackage[english]{babel}
\usepackage{siunitx}
\usepackage[T1]{fontenc}
\usepackage[utf8]{inputenc}
\usepackage{subfigure}
\usepackage{graphicx}
\usepackage{dcolumn}
\usepackage{bm}
\date{\today}
\title{}
\begin{document}

\keywords {
  Highly charged ions;
  Charge breeder;
  Charge breeding;
  EBIT;
  Ion traps;
  High-precision mass measurements;
  Penning trap;
  Exotic isotopes;
  Radioactive isotopes;
  mass spectrometry;
}

\title{Mass measurements of neutron-rich Rb and Sr isotopes}

\author{R. Klawitter}
\affiliation{TRIUMF, 4004 Wesbrook Mall, Vancouver BC, V6T 2A3 Canada}
\affiliation{Max Planck Institut f{\"u}r Kernphysik, 69117 Heidelberg, Germany}

\author{A. Bader}
\affiliation{TRIUMF, 4004 Wesbrook Mall, Vancouver BC, V6T 2A3 Canada}
\affiliation{{\'E}cole des Mines de Nantes, 4, rue Alfred Kastler, B.P. 20722, F-44307, Nantes, France}

\author{M. Brodeur}
\affiliation{Department of Physics, University of Notre Dame, Notre Dame, Indiana 46556 USA}

\author{U. Chowdhury}
\affiliation{Department of Physics, University of Manitoba, R3T 2N2, Winnipeg, Canada}

\author{A. Chaudhuri}
\affiliation{TRIUMF, 4004 Wesbrook Mall, Vancouver BC, V6T 2A3 Canada}

\author{J. Fallis}
\affiliation{TRIUMF, 4004 Wesbrook Mall, Vancouver BC, V6T 2A3 Canada}

\author{A.T. Gallant}
\affiliation{TRIUMF, 4004 Wesbrook Mall, Vancouver BC, V6T 2A3 Canada}
\affiliation{Department of Physics \& Astronomy, University of British Columbia, Vancouver BC, V6T 1Z1 Canada}

\author{A. Grossheim}
\affiliation{TRIUMF, 4004 Wesbrook Mall, Vancouver BC, V6T 2A3 Canada}

\author{A.A. Kwiatkowski}
\affiliation{Cyclotron Institute and Deptartment of Physics and Astronomy, Texas A\&M University, College Station, TX 77843, USA}
\affiliation{TRIUMF, 4004 Wesbrook Mall, Vancouver BC, V6T 2A3 Canada}

\author{D. Lascar}
\affiliation{TRIUMF, 4004 Wesbrook Mall, Vancouver BC, V6T 2A3 Canada}

\author{K. G. Leach}
\affiliation{TRIUMF, 4004 Wesbrook Mall, Vancouver BC, V6T 2A3 Canada}

\author{A. Lennarz}
\affiliation{TRIUMF, 4004 Wesbrook Mall, Vancouver BC, V6T 2A3 Canada}
\affiliation{Institut f{\"u}r Kernphysik, Westf{\"a}lische Wilhelms-Universit{\"a}t, 48149 M{\"u}nster, Germany}

\author{T.D. Macdonald}
\affiliation{TRIUMF, 4004 Wesbrook Mall, Vancouver BC, V6T 2A3 Canada}
\affiliation{Department of Physics \& Astronomy, University of British Columbia, Vancouver BC, V6T 1Z1 Canada}

\author{J. Pearkes}
\affiliation{Department of Physics \& Astronomy, University of British Columbia, Vancouver BC, V6T 1Z1 Canada}

\author{S. Seeraji}
\affiliation{Department of Chemistry, Simon Fraser University, Burnaby, British Columbia V5A 1S6, Canada}

\author{M.C. Simon}
\affiliation{Stefan Meyer Institute for Subatomic Physics, Austrian Academy of Sciences, 1090 Vienna, Austria}

\author{V.V. Simon}
\affiliation{Physikalisches Institut der Universit{\"a}t Heidelberg, 69120 Heidelberg, Germany}

\author{B.E. Schultz}
\affiliation{Department of Physics, University of Notre Dame, Notre Dame, Indiana 46556 USA}

\author{J. Dilling}
\affiliation{TRIUMF, 4004 Wesbrook Mall, Vancouver BC, V6T 2A3 Canada}
\affiliation{Department of Physics \& Astronomy, University of British Columbia, Vancouver BC, V6T 1Z1 Canada}

\begin{abstract}
  We report on the mass measurements of several neutron-rich
  $\mathrm{Rb}$
  and $\mathrm{Sr}$
  isotopes in the $A \approx 100$
  region with the TITAN Penning-trap mass spectrometer. Using highly
  charged ions in the charge state $q=10+$,
  the masses of $^{98,99}\mathrm{Rb}$
  and $^{98-100}\mathrm{Sr}$
  have been determined with a precision of $6 - 12\ \mathrm{keV}$,
  making their uncertainty negligible for \textit{r}-process
  nucleosynthesis network calculations. The mass of
  $^{101}\mathrm{Sr}$
  has been determined directly for the first time with a precision
  eight times higher than the previous indirect measurement and a
  deviation of $3\sigma$
  when compared to the Atomic Mass Evaluation.  We also confirm the
  mass of $^{100}\mathrm{Rb}$ from a previous
  measurement. 
  Furthermore, our data indicates the existance of a low-lying isomer
  with $80\ \mathrm{keV}$
  excitation energy in $^{98}\mathrm{Rb}$.
  We show that our updated mass values lead to minor changes in the
  \textit{r}-process by calculating fractional abundances in the $A\approx 100$
  region of the nuclear chart.
\end{abstract}

\maketitle

\section{Introduction}
\label{sec:orgheadline1}

Understanding the production of the heaviest elements found in nature
is one of the most challenging open question for all of physics
\cite{03:_connec_quark_cosmos}. Although it is generally understood that
half of the elements heavier than iron (Z > 26) are synthesized in a
series of rapid nuclear reactions and subsequent decays
(\textit{r}-process) in short, violent stellar explosions, there is no
consensus on the exact astrophysical events responsible
\cite{arnould07,thielemann11:_what}. The \textit{r}-process is fueled by
an extreme overabundance of neutrons (\(n_n > 10^{20}
\mathrm{cm^{-3}}\)) in a hot (\(T \gtrapprox 10^{9}\) K) environment,
creating very neutron-rich isotopes by sequential, rapid neutron
captures \((n,\gamma)\) which are eventually balanced by
photodisintegrations \((\gamma,n)\). Starting from a seed nuclide,
usually around iron, it creates a path through the isotopic
chart up to uranium, with \(\beta\)-decays connecting the isotopic
chains. When the temperature and neutron density drop, the process
"freezes out", and the neutron-rich isotopes \(\beta\)-decay to stability.

The path describing the \emph{r}-process through the nuclear chart strongly
depends on the neutron-separation energies (\(S_n\)) of nuclei on and
around it. Likely scenarios for the \emph{r}-process \cite{cowan91:_r} take
place in environments where neutron captures and photodisintegrations
are much faster than \(\beta\)-decays. Therefore, a \((n,\gamma)
\rightleftharpoons (\gamma,n)\) equilibrium is reached in each isotopic
chain, and the yields \(Y_i\) of neighboring isotopes satisfy the nuclear
Saha equation \cite{schatz13:_nuclear}:
\begin{equation}
  \label{eq:saha:fraq:abundances}
  \frac{Y_1}{Y_2} =
  n_n \frac{G_1}{2G_2}
    \left(
      \frac{A+1}{A} \frac{2\pi\hbar^2}{m_ukT}
    \right)^{3/2} e^{S_n/kT}
\end{equation}

for partition functions \(G_i\), Boltzmann constant \(k\), temperature
\(T\), atomic mass unit \(m_u\), and mass number \(A\). Most of the
abundance in a chain is carried by one or two "waiting-points"
\cite{burbidge57:_synth_elemen_stars} that inhibit capture to
heavier nuclei until their \(\beta\)-decay enables further, energetically
favorable neutron captures in the next isotopic chain.

Waiting points have a large impact on the final abundance patterns,
especially when their neutron number is magic (\(N = 50, 82, 126\)).  In
those cases, \(S_n\) can drop markedly, and the \emph{r}-process runs through
several, rather long \(\beta\)-decays \cite{brett12:_sensit}. \(S_{n}\) for
the most likely \emph{r}-process environment with temperatures of \(T \sim
10^{9}K\) and neutron densities of \(n_{n} > 10^{20}\mathrm{cm}^{-3}\)
are expected to be \(2-3\ \mathrm{MeV}\) \cite{cowan91:_r}. The path runs
therefore through very neutron-rich, unstable nuclei. These nuclides
are challenging to produce in the laboratory and to measure with the
required accuracy, as an uncertainty of \(10-100\ \mathrm{keV}\) on the
\(S_n\) is required to consider it negligible in \emph{r}-process models.  As a
consequence, only a few waiting-point masses have been derived
directly \cite{schatz13:_nuclear}. Where direct mass measurements are
not yet feasible, the \(S_n\) has to be extracted from theoretical
models
(e.g. \cite{moller95:_nuclear_groun_state_masses_defor,duflo95:_micros,pearson96:_nuclear_bogol}),
which can differ substantially in predicted mass values for nuclei far
from stability. One such region of waiting points is assumed to be at
\(A\approx 100,\ N\approx 60\) where rapid shape transitions are not
well reproduced by theoretical mass models
\cite{rodriguez-guzman10:_signat,simon12:_pennin_rb_sr}. Experimental
mass values for neutron-rich isotopes, especially in the vicinity of
\emph{r}-process waiting points, are crucial to further constrain
nucleosynthesis models as well as to help increase the predictive
power of mass models far from stability.

Ion traps are well established as the tool of choice for mass
measurements of short-lived isotopes
\cite{blaum13:_precis_atomic_physic_techn_nuclear}. TRIUMF's Ion Trap
for Atomic and Nuclear science (TITAN)
\cite{kwiatkowski14:_titan,dilling06:_mass_titan} is optimized to make
the fast, precision mass measurements, and it is the only facility of
its kind employing an electron beam ion trap (EBIT) to charge-change
short-lived isotopes for the pupose of Penning-trap mass spectrometry.
In this article we report on Penning-trap mass measurements of
\(^{98,99}\mathrm{Rb}\) and \(^{98-100}\mathrm{Sr}\) with TITAN and
show how the updated mass values change the fractional abundances
of the \emph{r}-process in the \(A\approx 100,\ N\approx 60\) region of the
nuclear chart.

\section{Experimental Method}
\label{sec:orgheadline2}

\begin{figure}[htb]
\centering
\includegraphics[width=0.5\textwidth]{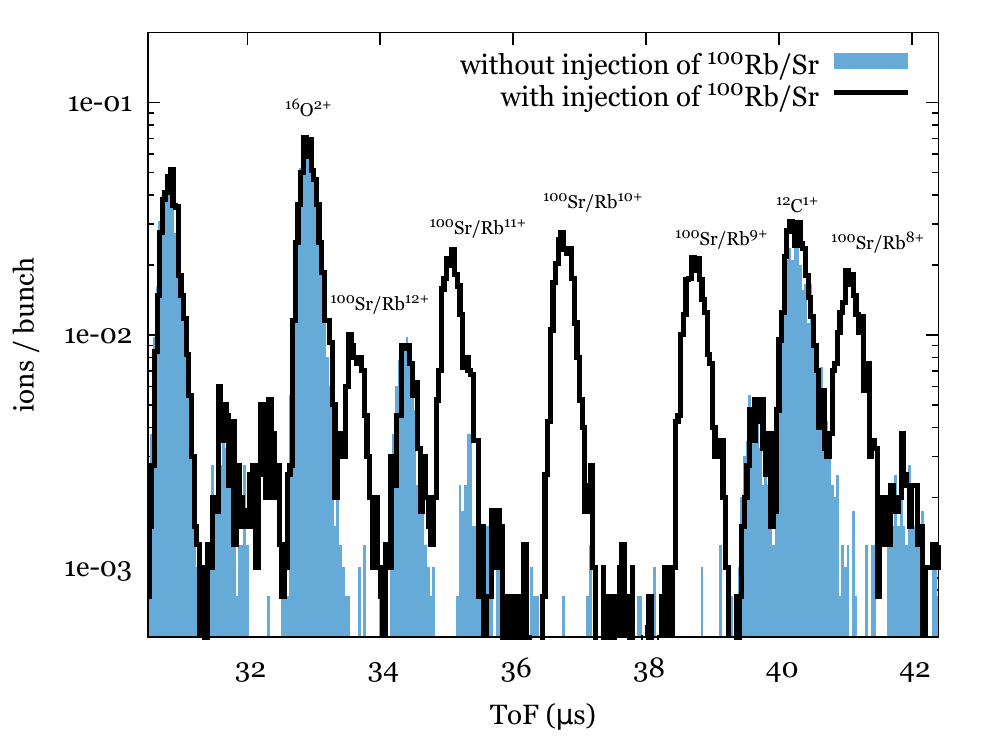}
\caption{\label{fig:orgparagraph1}
(color online) Typical time-of-flight (TOF) spectrum for ions extracted from the EBIT recorded with a micro-channel plate detector (MCP) just upstream of MPET. Charge state \(q=10+\) (central peak) was used for mass determination of all the isotopes to minimize the amount of ionized residual gas (colored peaks) entering the Penning trap.}
\end{figure}

TITAN is coupled to the Isotope Separator and Accelerator (ISAC)
\cite{dilling14:_isac} facility at TRIUMF, where, for the experiment
described, \(8\ \mathrm{\mu A}\) of \(480\ \mathrm{MeV}\) protons were
directed onto a uranium carbide target \cite{kunz13:_compos_trium} to
produce radioactive isotopes. These were then ionized using a surface
ion source, extracted, and accelerated to \(20\ \mathrm{keV}\) beam
energy. The neutron-rich Rb/Sr isotopes of interest were subsequently
separated by a dipole magnet with a resolving power of up to \(m/\delta
m \approx 3000\) \cite{dilling14:_rare_isac}. After selection the beam of
interest was directed to and injected into the TITAN Radio Frequency
Quadrupole (RFQ, \cite{brunner12:_titan_rfq,smith06:_first_titan_rfq})
cooler and buncher.  Here, the accumulated beam was cooled by
interacting with a helium buffer gas for a few ten to a few hundred
milliseconds, extracted in bunches, and directed towards the electron beam
ion trap (EBIT, \cite{lapierre10:_titan_ebit}).

\(\mathrm{Rb/Sr}\) bunches captured in the EBIT were collisionally
ionized further by an electron beam with current \(I_e = 100\ \mathrm{mA}\) and
energy \(E_e = 2.5\ \mathrm{keV}\), resulting in a distribution of
charge states (see Fig. \ref{fig:orgparagraph1}).
The \(q=10+\) charge state was then selected by time of flight (TOF)
with a Bradbury-Nielson gate (BNG, \cite{brunner12:_bradb_niels}).

After being injected into the measurement Penning trap (MPET), the
cyclotron frequency \(\nu_c\) of the ion of interest was measured using
the time-of-flight ion-cyclotron-resonance technique (TOF-ICR,
\cite{koenig95:_quadr}, see also Fig. \ref{fig:orgparagraph2}). The mass \(m^+\) of
an ion in a homogeneous magnetic field with strength \(B\) can then be
inferred from:

\begin{equation}
\label{eq:cyclfreq}
  2\pi\nu_c = \frac{q\cdot e}{m^+} B,
\end{equation}

with \(q\) being the ion charge state and \(e\) the elementary charge. The
precision is given by:
\begin{equation}
  \label{eq:tof-icr-precision}
\frac{\Delta m}{m} = F \cdot
\frac{m}{qeBT_{RF}\sqrt{N_{ion}}},
\end{equation}

where \(T_{RF}\) the excitation time, \(N_{ion}\) the number of observed
ions, and \(F\) an apparatus-dependent proportionality constant close to
unity for TITAN \cite{brodeur12:_verif_titan_pennin}. The excitation
time is restricted by the ion's half-life; the magnetic field is
limited technically to \(4-10\ \mathrm{T}\); and, \(N_{ion}\) depends on
the isotopes' yield and the overall system efficiency.  With an EBIT,
however, the charge state of radioactive ions can be increased within
a few milliseconds, providing a powerful tool for high-precision mass
measurements on short-lived isotopes far from stability
\cite{ettenauer11:_first_use_high_charg_states,malbrunot-ettenauer15:_pennin,frekers13:_pennin_q}.
TITAN is currently the only facility in the world using highly charged
ions (HCI) for the purpose of Penning-trap mass spectrometry (PTMS) on
short-lived isotopes, bringing unique opportunities and challenges
\cite{ettenauer13:_advan_pennin}.  Since yields for isotopes far from
stability are generally low, with rates down to a few ions per second,
efficient ion trapping, extraction, and transport become still more
important. For the experiment described in this work, total system
efficiency was around \(0.1\%\), limiting measurements to isotopes with
yields from ISAC of about \(10^{3}\ \mathrm{ions}/s\).
\section{Analysis and Results}
\label{sec:orgheadline10}
\begin{figure}[htb]
\centering
\includegraphics[width=0.5\textwidth]{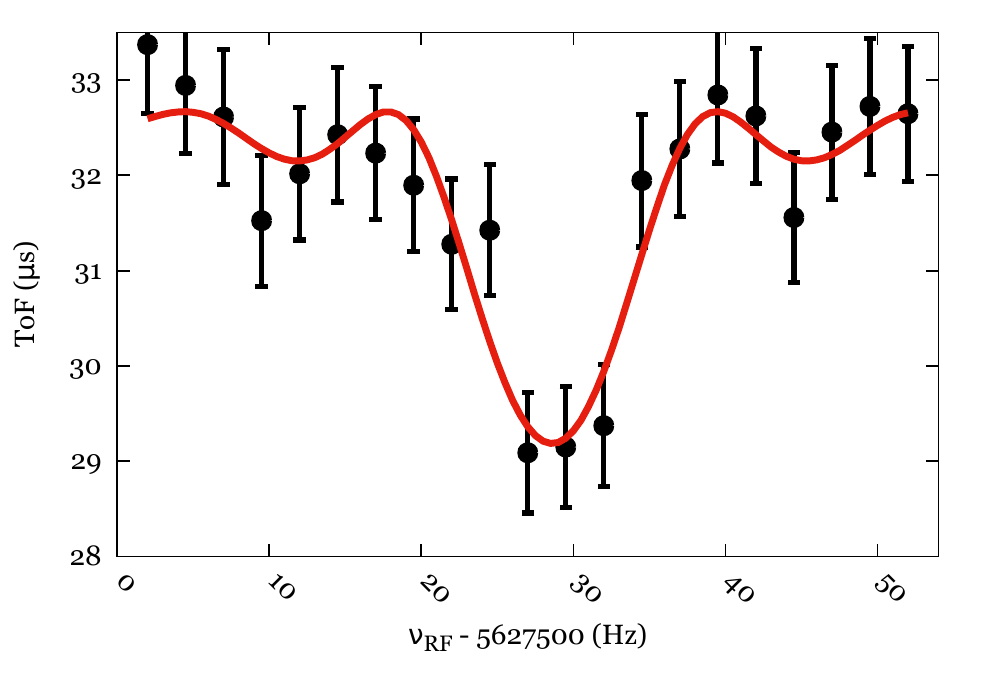}
\caption{\label{fig:orgparagraph2}
(color online) Time-of-flight (TOF) resonance of \(^{101}\mathrm{Sr}\) [\(T_{1/2} = 118(3)\ m\mathrm{s}\)]. Black dots: average time-of-flight and associated standard deviation; continous red line: fit \cite{koenig95:_quadr} to the data.}
\end{figure}

To remove the dependency on the magnetic field in eq. \ref{eq:cyclfreq}, the ratio
of two cyclotron frequencies:

\begin{equation}
\label{eq:freq-ratios}
R = \frac{v_{c,ref}}{v_c} = \frac{q_{ref}\cdot m^+}{q\cdot m_{ref}^+}
\end{equation}
is recorded, where \(q, q_{ref}\) are the charge state and \(m^+, m^+_{ref}\) the
mass of the ion of interest and of a well-known reference ion respectively. To
compute \(R\), we extracted the cyclotron frequency of the ion of interest and of
the reference ion from a fit of a well-known analytic form \cite{koenig95:_quadr}
to the data; see Fig. \ref{fig:orgparagraph2} for an example. The atomic mass \(m\) of
the isotope can then be calculated from the ratio \(R\) using:
\begin{equation}
  \label{eq:mass-from-ratio}
  m = \frac{q}{q_{ref}\cdot R}\cdot
  \left(  m_{ref} - q_{ref}\cdot m_e + B_{e,ref} \right) + q\cdot m_e - B_e,
\end{equation}

where \(B_e, B_{e,ref}\) are the atomic binding energy of the ion of interest
and reference ion respectively, and \(m_e\) is the mass of the electron.

A number of systematic effects can influence the result and have to be
taken into account.  Although we selected a specific charge state to
minimize background with the BNG, ion bunches from the EBIT contain
contaminant ions that will enter the Penning trap (see Fig.
\ref{fig:orgparagraph1}). Furthermore, HCI can charge exchange with neutral
background atoms, creating ions of different charge states inside the
trap.

We minimized the effects of contaminants and of more than one trapped
ion \cite{kellerbauer03:_from} in a number of ways.  Firstly, a 10 ms
long dipole cleaning excitation \cite{blaum04:_popul_pennin} was used to
remove known contaminants from the trap. Secondly, the beam transfer
rate was adjusted to have, on average, less than one detected ion per
measurement cycle. Thirdly, all values are based on an analysis that
includes only cycles with one detected ion. The difference in \(R\)
using all cycles to fit the data as compared to only considering
cycles with one detected ion was added as a systematic uncertainty
\cite{ettenauer11:_first_use_high_charg_states}. As the number of total
detected ions was low, especially for the shortest-lived isotopes
investigated, ``count-class''-type analysis \cite{kellerbauer03:_from}
could not be used. However, results for species with more than three
detected ions per cycle on average were cross-checked using a
count-class analysis, showing agreement with the non-count-class
analysis to within \(0.3\sigma\).  Finally, for ion events to be
included in the analysis, the ion time of flight had to be within a
certain range, \(20\ \mathrm{\mu s} \leq t \leq 50\ \mathrm{\mu
s}\). This excludes events that are due to light charge-exchange
products (e.g. \(\mathrm{H^{+}_{2}}\)) as well as limits the effect of
other contaminants. It was found in an off-line study, that
contaminants typically populate the TOF spectrum in this range and can
be excluded in this way.  However, as the correct range cannot be
selected unequivocally, the maximal shift in \(R\) due to a set of
different TOF ranges (\(20\ \mathrm{\mu s} \leq t \leq 40\ \mathrm{\mu
s}\), \(20\ \mathrm{\mu s} \leq t \leq 60\ \mathrm{\mu s}\)) was added as a
systematic uncertainty.

Since the described experiment was performed with HCI, the difference in atomic
binding energies of reference ion and ion of interest is on the order of a few
hundred eV and has to be taken into account.  We used binding energies from
\cite{rodrigues04:_system}, with an estimated total uncertainty of \(5\ \mathrm{eV}\)
\cite{simon12:_pennin_rb_sr}, which can be neglected.

At TITAN, reference ions and ions of interest cannot be measured
simultaneously. The reference mass is therefore interpolated linearly
from two reference measurements (\(^{85}\mathrm{Rb}\) in our case), one
taken before and one taken after the measurement of the ion of
interest. As eq.~\ref{eq:mass-from-ratio} is valid only for a
constant magnetic field, fluctuations over time in the magnetic field
will affect the determined mass. For our setup these fluctuations were
shown to be below \(\Delta R/R = 0.2 \ \mathrm{ppb / h}\)
\cite{brodeur09:_new_pennin_titan}. To reduce their effect the period
for a single measurement was limited to \(0.5\ \mathrm{h}\). The mass of
\(^{85}\mathrm{Rb}\) is known to within \(5\ \mathrm{eV}\)
\cite{wang12:_ame20} and therefore also negligible.

Systematic uncertainties due to relativistic effects, field
inhomogeneities, distortions or misalignment are either proportional
to the difference in \({q}/{m}\) or \({m}/{q}\) of ion of interest and
reference ion. To minimize any such effect, we choose a reference
\(m/q\) as close as possible to the \(m/q\) of the species of interest,
with \(1.038 < \frac{m \cdot q_{\mathrm{ref}}}{q \cdot
m_{\mathrm{ref}}} < 1.070\) by preparing stable \(^{85}\mathrm{Rb}\) from
the TITAN off-line source \cite{kwiatkowski14:_titan} in charge state
\(q=9+\), where \(m\) is given in atomic mass units.  Spatial magnetic
field inhomogeneities and harmonic distortions as well as misalignment
of the magnetic field axis are known to be \(\Delta R/R=3\cdot
10^{-11}\), and \(\Delta R/R=3\cdot 10^{-10}\)
\cite{brodeur09:_new_pennin_titan} respectively and can therefore be
neglected as well.

Since HCI experience larger velocities inside the Penning trap than
SCI, relativistic effects can become important and have to be
considered. Following \cite{brodeur09:_new_pennin_titan} their influence
is below \(\Delta R/R = 9\cdot 10^{-10}\) for the results presented and
have not been included in the reported uncertainties. To exclude other
systematic effects and monitor system performance, we made frequent
measurements of the mass of \(^{85}\mathrm{Rb}^{8+}\) with
\(^{85}\mathrm{Rb}^{9+}\) as reference ions, showing good agreement
(\(\delta m\) below \(0.1\sigma\)) with the literature value
\cite{wang12:_ame20}.

The frequency ratios of all measured isotopes relative to
\(^{85}\mathrm{Rb}^{9+}\) are summarized in Table \ref{tab:results},
together with the corresponding mass excess and, for comparison, mass
excess values found in the literature.

\begin{table*}\label{tab:results}\def\arraystretch{1.4}\setlength{\tabcolsep}{0.2cm}
\begin{center}
\caption{Frequency ratios of $^{98-100}\mathrm{Rb}^{10+}$ and $^{98-101}\mathrm{Sr}^{10+}$ isotopes relative to $^{85}\mathrm{Rb}^{9+}$ as well as atomic mass excesses. The first uncertainty on the frequency ratio represents the statistical uncertainty multiplied by the reduced-$\chi^2$ of the fit to the data. The second and third uncertainties reflect ambiguities in the choice of time-of-flight range and number of detected ions included in the resonance \cite{ettenauer11:_first_use_high_charg_states} respectively. The fourth uncertainty in curly brackets represents the quadrature sum of all the uncertainties. For the mass excesses the combined uncertainty is shown. For comparison the last three columns contain mass excess values as determined by ISOLTRAP \cite{manea13:_collec_rb}, JYFLTRAP \cite{hager06:_first_precis_mass_measur_refrac_fission_fragm}, and the values quoted in the AME 2012 \cite{wang12:_ame20} . \vspace{0.5em}}
\begin{tabular}{rllllllll}
\hline
\hline
 & \(T_{1/2}\) & \# ions & \(T_{rf}\) & Frequency ratio \(r = \nu_{ref}^{+ }/\nu\) & ME & ME\(_{\text{ISOL}}\) & ME\(_{\text{JYFL}}\) & ME\(_{\text{AME}}\)\\
 & \((\mathrm{ms})\) &  & (ms) &  & (keV) & (keV) & (keV) & (keV)\\
\hline
\(^{98}\mathrm{Rb}\) & 114(5) & 5737 & 80 & 1.038109144(48)(39)(12)\{63\} & -54319.6(5.5) & -54309.4(4.0) &  & -54318.3(3.4)\\
\(^{99}\mathrm{Rb}\) & 54(4) & 5095 & 80 & 1.04874535(07)(08)(03)\{11\} & -51124.6(9.3) & -51120.3(4.5) &  & -51205(112)\\
\(^{100}\mathrm{Rb}\) & 51(4) & 482 & 30 & 1.0594014(09)(11)(06)\{15\} & -46190(140) & -46247(20) &  & -46547(196)\\
\(^{98}\mathrm{Sr}\) & 653(4) & 6346 & 80 & 1.037971485(38)(59)(30)\{76\} & -66416.6(6.7) &  & -66431(10) & -66426.0(3.7)\\
\(^{99}\mathrm{Sr}\) & 269(1) & 5270 & 80 & 1.048615651(46)(54)(21)\{74\} & -62522.4(6.5) &  & -62524(7) & -62511.9(3.6)\\
\(^{100}\mathrm{Sr}\) & 202(4) & 7110 & 80 & 1.05924630(11)(06)(04)\{13\} & -59816(11) &  & -59828(10) & -59830.1(9.5)\\
\(^{101}\mathrm{Sr}\) & 118(3) & 2760 & 80 & 1.06989723(11)(02)(01)\{11\} & -55327.6(9.8) &  &  & -55562(81)\\
\hline
\hline
\end{tabular}
\end{center}
\end{table*}

\subsection{\(^{98}\mathrm{Rb}\) (ground and isomeric state)}
\label{sec:orgheadline3}
We determined the mass excess of the produced state of
\(^{98}\mathrm{Rb}\) to be \(-54319.6(5.5)\mathrm{keV}\), in agreement
(\(0.22\sigma\)) with a previous TITAN mass measurement and the Atomic
Mass Evaluation (AME 2012) value, but disagreeing (\(1.85\sigma\)) with a
recent determination by ISOLTRAP \cite{manea13:_collec_rb}.

In an effort to identify a proposed low-lying isomeric state
\cite{lhersonneau02:_k}, with an estimated excitation energy of \(\approx
280(128)\ \mathrm{keV}\) \cite{wang12:_ame20} and \(T_{1/2}=96(3)\
\mathrm{ms}\), we searched for a characteristic signature
\cite{gallant12:_highl_pennin} within a range of \(\pm 2.5\ \mathrm{MeV}\)
around the known state. However, the observed time-of-flight
distribution for ions in resonance is consistent with a predominantly
populated very low lying isomeric state of \(^{98}\mathrm{Rb}^m\) and an
admixture of about 30(10)\% of the \(^{98}\mathrm{Rb}^g\) ground
state. Using a double-resonance \cite{gallant12:_highl_pennin} function
to fit the data yields an excitation energy of \(80\ \mathrm{keV}\) and
leads to a decrease in the reduced-\(\chi^2\) of the fit by about 15\%.
Due to insufficient resolution the associated uncertainty is about as
big as the excitation energy itself.

\begin{figure}[htb]
\centering
\includegraphics[width=0.5\textwidth]{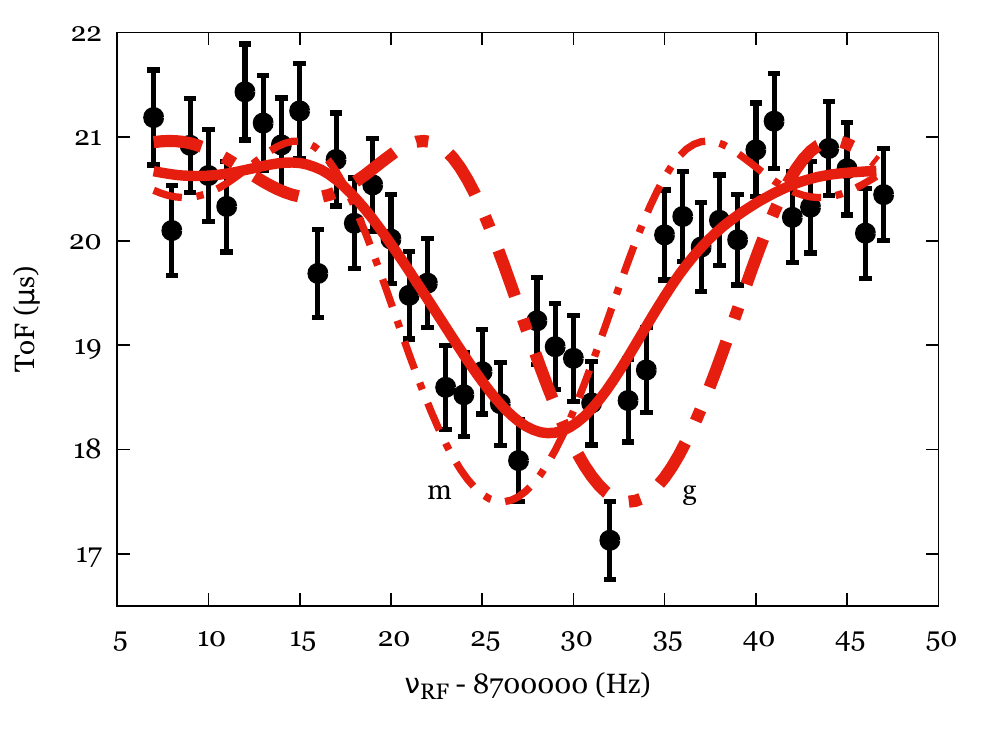}
\caption{\label{fig:orgparagraph3}
(color online) Black points: All TOF-ICR data for \(^{98}\mathrm{Rb^{g,m;q=15+}}\) from \cite{simon12:_pennin_rb_sr} added. Dashed lines: Individual contributions of \(^{98}\mathrm{Rb^{m;q=15+}}\) and \(^{98}\mathrm{Rb^{g;q=15+}}\) respectively as well as fit to both states together (solid line). The reduced-\(\chi^2\) for a two-state model is \(\chi^2_{r,g+m} = 1.7\) as compared to \(\chi^2_r = 2\) for a one-state model.}
\end{figure}

The discussion about a long-lived isomeric state was recently renewed
by results from a collinear laser spectroscopy experiment on
\(^{98}\mathrm{Rb}\) \cite{procter15:_direc}, that used the identical
production mechanism. With this in mind we reanalyzed all data from
our 2012 publication \cite{simon12:_pennin_rb_sr}. The previous
experiment used \(^{98}\mathrm{Rb}\) in charge state \(q=15\) instead of
\(q=10\), giving correspondingly better resolving power. We extracted
the mass excesses from a double resonance fit (see figure
\ref{fig:orgparagraph3}), calculating \(-54296(20)\) keV for the isomeric state
and \(-54376(30)\) keV for the ground state.  The relative populations
of 35(15)\% to 65(15)\%, respectively, are in agreement with the values
obtained in the collinear laser spectroscopy expermiment
\cite{procter15:_direc}.

\subsection{\(^{99}\mathrm{Rb}\)}
\label{sec:orgheadline4}
We measured the mass excess of \(^{99}\mathrm{Rb}\ [T_{1/2}=54(4)
\mathrm{ms}]\) to be \(-51124.6(9.3)\ \mathrm{keV}\), which is in
agreement with a recent ISOLTRAP measurement by Manea et
al. \cite{manea13:_collec_rb}. Both values agree within \(1\sigma\) of the
value published in the AME 2012 \cite{wang12:_ame20,audi12:_ame20} but
carry less than a tenth of the uncertainty. The AME 2012 value was adopted
from a measurement employing a double-focusing mass spectrometer
\cite{audi86:_mass_rb_cs_fr} (with 13\% weight) and a determination using
the \(\beta\)-endpoint endpoint energy of \(^{99}\mathrm{Rb}\left(
\beta^- \right){^{99}\mathrm{Sr}}\) \cite{iafigliola84:_beta_rb_sr} (87\%
weight).

\subsection{\(^{100}\mathrm{Rb}\)}
\label{sec:orgheadline5}
We confirm the first PTMS measurement of \(^{100}\mathrm{Rb}\ [T_{1/2}=51(4)]\) by Manea
et al. \cite{manea13:_collec_rb}. Since the amount of
\(^{100}\mathrm{Sr}\) in the \(A=100\) beam delivered was more than an
order of magnitude higher than the amount of \(^{100}\mathrm{Rb}\), a
dipole RF cleaning period of \(10\ \mathrm{ms}\) was used reduce the
amount of isobaric contamination. The comparatively large uncertainty of
about \(140\ \mathrm{keV}\) for our value can therefore be explained by a
combination of the preperation time, a poor total system efficiency
and, as a consequence of both, the reduction in observed
events. Nevertheless, the measurement of \(^{100}\mathrm{Rb}\) sets a
new record for the shortest-lived isotope to be measured in a Penning
trap using HCI.

\subsection{\(^{98}\mathrm{Sr}\)}
\label{sec:orgheadline6}
For the Penning-trap mass measurement of \(^{98}\mathrm{Sr},\ \left[
T_{1/2}=653(4)\ \mathrm{ms}\right]\) we find a value of \(-62522.4(6.5)
\mathrm{keV}\) for the mass excess. This is within \(2\sigma\) of the
value found by a previous TITAN campaign \cite{simon12:_pennin_rb_sr}
(87\% of the AME 2012 \cite{wang12:_ame20,audi12:_ame20} value) and
within \(1.5\sigma\) of the value published by JYFLTRAP
\cite{hager06:_first_precis_mass_measur_refrac_fission_fragm} (13\% of
the AME 2012 value). We reanalyzed the value published by TITAN by
perfoming a second, independent data analysis, indicating that the
deviation is statistical in nature.

\subsection{\(^{99}\mathrm{Sr}\)}
\label{sec:orgheadline7}
For the mass excess of \(^{99}\mathrm{Sr}\ [T_{1/2}=269(1)\
\mathrm{ms}]\) we find a value of \(-66416.6(6.7) \mathrm{keV}\), which
is in agreement with a measurement performed by JYFLTRAP
(giving a weight of 24\% to the AME 2012 value) in
2006 \cite{hager06:_first_precis_mass_measur_refrac_fission_fragm}, but
agrees only within \(2.5\sigma\) with a recent TITAN measurement
\cite{simon12:_pennin_rb_sr} (76\% of the AME 2012 \cite{wang12:_ame20,audi12:_ame20} value).

We have reanalyzed the data taken during the first TITAN \(A\approx
100\ \mathrm{Rb/Sr}\) campaign, arriving at the same mass excess but at
a considerably larger uncertainty with \(\mathrm{ME\left( ^{99}Sr
\right)} = -62505(19)\ \mathrm{keV}\). This brings the three existing
Penning trap mesaurements into agreement.

\subsection{\(^{100}\mathrm{Sr}\)}
\label{sec:orgheadline8}
For \(^{100}\mathrm{Sr}\ [T_{1/2}=202(4)\ \mathrm{ms}]\) we found a mass
excess value of \(-59816(12)\ \mathrm{keV}\), in agreement with the only
Penning-trap mass measurement value published to date
\cite{hager06:_first_precis_mass_measur_refrac_fission_fragm} from
JYFLTRAP.

\subsection{\(^{101}\mathrm{Sr}\)}
\label{sec:orgheadline9}
We have carried out the first direct mass measurement (see
Fig. \ref{fig:orgparagraph2}) of \(^{101}\mathrm{Sr}\ \left[T_{1/2}=118(3)\
\mathrm{ms} \right]\) and report a mass excess value of \(-55327.6(9.8)
\mathrm{keV}\). This value marks a \(3\sigma\) deviation from the one
adopted in AME 2012 \cite{wang12:_ame20,audi12:_ame20}, which is based
on the result of a \(\beta\)-endpoint measurement \cite{balog92:_exper}
for \(^{101}\mathrm{Sr}\left( \beta^- \right){^{101}\mathrm{Y}}\). This
discrepancy between \(\beta\)-endpoint measurements and PTMS has been
documented for neutron-rich isotopes
\cite{van12:_mass_canad_pennin_trap}.

\section{Astrophysical impact}
\label{sec:orgheadline11}
\begin{figure}[htb]
\centering
\includegraphics[width=0.5\textwidth]{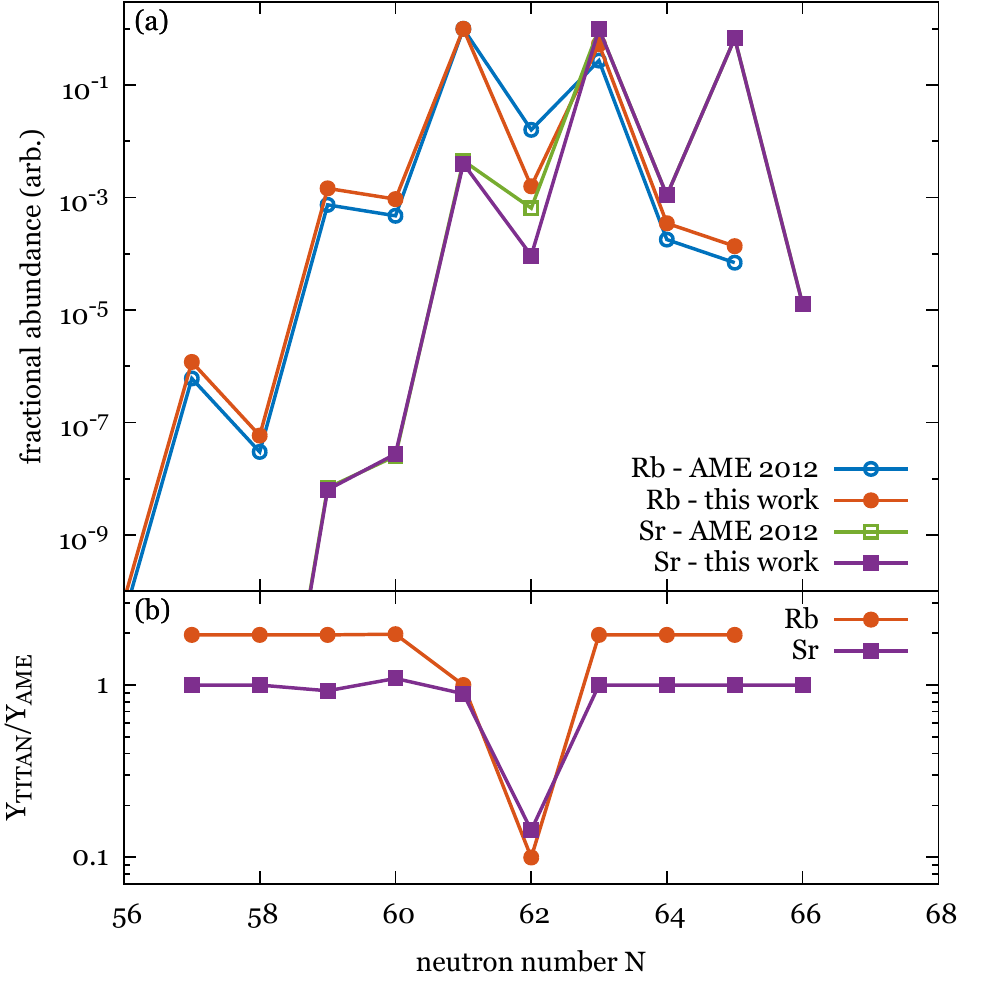}
\caption{\label{fig:orgparagraph4}
(color online) (a) Fractional \emph{r}-process abundances, each relative to the most abundant isotope, using the waiting-point approximation for Rb, Sr isotopic-chains with temperature \(T=1.4\cdot 10^9\) K and neutron density \(n_n=10^{20} \mathrm{cm^{-3}}\) for \(S_n\) from the AME 2012 \cite{wang12:_ame20,audi12:_ame20} (squares) and \(S_n\) from this work (circles). (b) The ratio \(Y_{TITAN}/Y_{AME 2012}\) corresponding to the fractional abundances shown in (a).}
\end{figure}

To illustrate the impact of the new data on the \emph{r}-process in the \(A
\approx 100\) region of the nuclear chart, we used
eq. \ref{eq:saha:fraq:abundances} to calculate fractional abundances
in the waiting-point approximation with temperatures \(10^8\ \mathrm{K}
\leq T \leq 10^{10}\ \mathrm{K}\), neutron densities \(10^{18}\
\mathrm{cm^{-3}} \leq n_n \leq 10^{25}\ \mathrm{cm^{-3}}\) and \(S_n\)
using data from this work and for comparison \(S_n\) from AME 2012
\cite{wang12:_ame20,audi12:_ame20}. Values for the partition functions
\(G_i\) were taken from, or interpolated using a spline function fitted
to the data given in, Rauscher et
al. \cite{rauscher00:_astrop_react_rates_from_statis_model_calcul}. The
biggest differences were found for astrophysical conditions with
\(n_{n} = 10^{20}\ \mathrm{cm^{-3}}\), \(T = 1.4\cdot 10^9 \mathrm{K}\),
with waiting points in the \(A\approx 100\) region, as illustrated in
Fig. \ref{fig:orgparagraph4}.

While the fractional abundances for Rb and Sr with \(N=62\) does change
by up to an order of magnitude, the new data do not change the
abundance pattern, i.e. location of the waiting points, significantly.

\section{Summary and future improvements}
\label{sec:orgheadline12}
\begin{figure}[htb]
\centering
\includegraphics[width=0.5\textwidth]{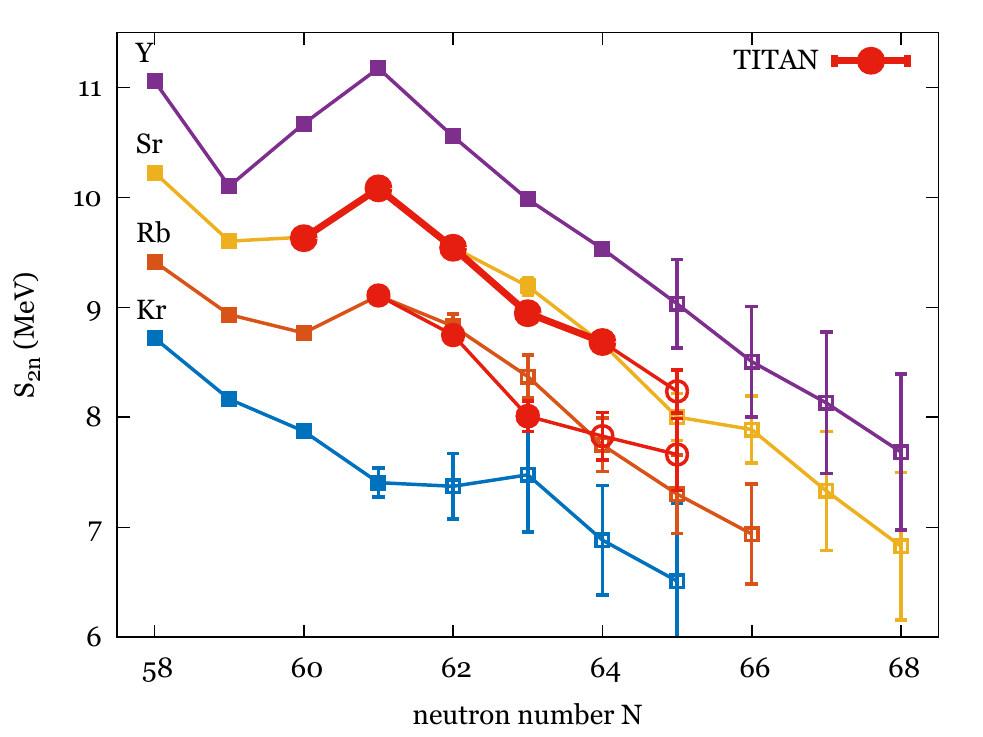}
\caption{\label{fig:orgparagraph5}
(color online) Two-neutron separation energies for \(Z = 36-39\) (Kr to Y) versus neutron number.  For comparison, \(S_{2n}\) based on new TITAN masses are plotted in red circles. \(S_{2n}\) graphed with open circles or squares are based on estimated mass values \cite{wang12:_ame20,audi12:_ame20}.}
\end{figure}

We have measured the masses of \(^{98,99}\mathrm{Rb}\)
and \(^{98-101}\mathrm{Sr}\) to a precision that makes mass
uncertainties negligible for \emph{r}-process models. The new data do not
change the fractional \emph{r}-process abundance pattern in the \(A=100\)
region significantly.  We have also presented first mass-spectrometric
evidence for a very low-lying isomer in \(^{98}\mathrm{Rb}\). The mass
of \(^{101}\mathrm{Sr}\) was determined directly for the first time, and
the mass determination of \(^{99,100}\mathrm{Rb}\) mark a new record for
the shortest-lived isotope measured in a Penning trap with HCI. For
illustration, the isotopic two-neutron separation energies (\(S_{2n}\))
are plotted in Fig. \ref{fig:orgparagraph5} with data from this work and from the
Atomic Mass Evaluation 2012 \cite{wang12:_ame20,audi12:_ame20}.

Mass measurements of isotopes even further from stability were
prohibited by a combination of current system efficiency and
significant isobaric contamination.  Both challenges will be met with
the addition of two new traps to the current setup in the near
future. A cooler Penning trap (CPET, \cite{ke06:_titan_trium}) will be
installed to reduce energy spread and therefore increase capture
efficiency as well as signal strength in MPET. To suppress isobaric
contamination, a multi-reflection time of flight spectrometer (MR-TOF,
\cite{jesch15:_mr_tof_ms_titan_trium}) will be installed downstream of
the RFQ cooler and buncher with a mass resolving power of up to
\(m/\Delta m = 50000\). Together these measures should enable TITAN to
determine masses of all the neutron-rich Sr and Rb isotopes important
for the \emph{r}-process to a precision, that makes mass-uncertainties
neglible for \emph{r}-process models.

\begin{acknowledgements}
The authors express their gratitude to the TRIUMF technical staff, the
ISAC beam delivery group, and M. Good for their contributions in
realizing this experiment. This work has been supported by the Natural
Sciences and Engineering Research Council of Canada (NSERC) and the
National Research Council of Canada. The authors acknowledge financial
support for T.D.M. from the NSERC CGS-M program, A.T.G. from the NSERC
CGS-D, A.L. from the Deutsche Forschungsgemeinschaft (DFG) under grant
FR 601/3-1, and V.V.S. from the Studienstiftung des Deutschen Volkes.
\end{acknowledgements}

\vfill

\bibliography{neutron_rich_rb_sr}

\begin{thebibliography}{46}%
\makeatletter
\providecommand \@ifxundefined [1]{%
 \@ifx{#1\undefined}
}%
\providecommand \@ifnum [1]{%
 \ifnum #1\expandafter \@firstoftwo
 \else \expandafter \@secondoftwo
 \fi
}%
\providecommand \@ifx [1]{%
 \ifx #1\expandafter \@firstoftwo
 \else \expandafter \@secondoftwo
 \fi
}%
\providecommand \natexlab [1]{#1}%
\providecommand \enquote  [1]{``#1''}%
\providecommand \bibnamefont  [1]{#1}%
\providecommand \bibfnamefont [1]{#1}%
\providecommand \citenamefont [1]{#1}%
\providecommand \href@noop [0]{\@secondoftwo}%
\providecommand \href [0]{\begingroup \@sanitize@url \@href}%
\providecommand \@href[1]{\@@startlink{#1}\@@href}%
\providecommand \@@href[1]{\endgroup#1\@@endlink}%
\providecommand \@sanitize@url [0]{\catcode `\\12\catcode `\$12\catcode
  `\&12\catcode `\#12\catcode `\^12\catcode `\_12\catcode `\%12\relax}%
\providecommand \@@startlink[1]{}%
\providecommand \@@endlink[0]{}%
\providecommand \url  [0]{\begingroup\@sanitize@url \@url }%
\providecommand \@url [1]{\endgroup\@href {#1}{\urlprefix }}%
\providecommand \urlprefix  [0]{URL }%
\providecommand \Eprint [0]{\href }%
\providecommand \doibase [0]{http://dx.doi.org/}%
\providecommand \selectlanguage [0]{\@gobble}%
\providecommand \bibinfo  [0]{\@secondoftwo}%
\providecommand \bibfield  [0]{\@secondoftwo}%
\providecommand \translation [1]{[#1]}%
\providecommand \BibitemOpen [0]{}%
\providecommand \bibitemStop [0]{}%
\providecommand \bibitemNoStop [0]{.\EOS\space}%
\providecommand \EOS [0]{\spacefactor3000\relax}%
\providecommand \BibitemShut  [1]{\csname bibitem#1\endcsname}%
\let\auto@bib@innerbib\@empty
\bibitem [{03:(2003)}]{03:_connec_quark_cosmos}%
  \BibitemOpen
  \href
  {http://www.nap.edu/catalog/10079/connecting-quarks-with-the-cosmos-eleven-science-questions-for-the}
  {\emph {\bibinfo {title} {Connecting Quarks with the Cosmos: Eleven Science
  Questions for the New Century}}}\ (\bibinfo  {publisher} {The National
  Academies Press},\ \bibinfo {address} {Washington, DC},\ \bibinfo {year}
  {2003})\BibitemShut {NoStop}%
\bibitem [{\citenamefont {Arnould}\ \emph {et~al.}(2007)\citenamefont
  {Arnould}, \citenamefont {Goriely},\ and\ \citenamefont
  {Takahashi}}]{arnould07}%
  \BibitemOpen
  \bibfield  {author} {\bibinfo {author} {\bibfnamefont {M.}~\bibnamefont
  {Arnould}}, \bibinfo {author} {\bibfnamefont {S.}~\bibnamefont {Goriely}}, \
  and\ \bibinfo {author} {\bibfnamefont {K.}~\bibnamefont {Takahashi}},\ }\href
  {\doibase 10.1016/j.physrep.2007.06.002} {\bibfield  {journal} {\bibinfo
  {journal} {Physics Reports}\ }\textbf {\bibinfo {volume} {450}},\ \bibinfo
  {pages} {97 } (\bibinfo {year} {2007})}\BibitemShut {NoStop}%
\bibitem [{\citenamefont {Thielemann}\ \emph {et~al.}(2011)\citenamefont
  {Thielemann}, \citenamefont {Arcones}, \citenamefont {K{\"a}ppeli},
  \citenamefont {Liebend{\"o}rfer}, \citenamefont {Rauscher}, \citenamefont
  {Winteler}, \citenamefont {Fr{\"o}hlich}, \citenamefont {Dillmann},
  \citenamefont {Fischer}, \citenamefont {Martinez-Pinedo}, \citenamefont
  {Langanke}, \citenamefont {Farouqi}, \citenamefont {Kratz}, \citenamefont
  {Panov},\ and\ \citenamefont {Korneev}}]{thielemann11:_what}%
  \BibitemOpen
  \bibfield  {author} {\bibinfo {author} {\bibfnamefont {F.-K.}\ \bibnamefont
  {Thielemann}}, \bibinfo {author} {\bibfnamefont {A.}~\bibnamefont {Arcones}},
  \bibinfo {author} {\bibfnamefont {R.}~\bibnamefont {K{\"a}ppeli}}, \bibinfo
  {author} {\bibfnamefont {M.}~\bibnamefont {Liebend{\"o}rfer}}, \bibinfo
  {author} {\bibfnamefont {T.}~\bibnamefont {Rauscher}}, \bibinfo {author}
  {\bibfnamefont {C.}~\bibnamefont {Winteler}}, \bibinfo {author}
  {\bibfnamefont {C.}~\bibnamefont {Fr{\"o}hlich}}, \bibinfo {author}
  {\bibfnamefont {I.}~\bibnamefont {Dillmann}}, \bibinfo {author}
  {\bibfnamefont {T.}~\bibnamefont {Fischer}}, \bibinfo {author} {\bibfnamefont
  {G.}~\bibnamefont {Martinez-Pinedo}}, \bibinfo {author} {\bibfnamefont
  {K.}~\bibnamefont {Langanke}}, \bibinfo {author} {\bibfnamefont
  {K.}~\bibnamefont {Farouqi}}, \bibinfo {author} {\bibfnamefont {K.-L.}\
  \bibnamefont {Kratz}}, \bibinfo {author} {\bibfnamefont {I.}~\bibnamefont
  {Panov}}, \ and\ \bibinfo {author} {\bibfnamefont {I.}~\bibnamefont
  {Korneev}},\ }\href {\doibase 10.1016/j.ppnp.2011.01.032} {\bibfield
  {journal} {\bibinfo  {journal} {Progress in Particle and Nuclear Physics}\
  }\textbf {\bibinfo {volume} {66}},\ \bibinfo {pages} {346 } (\bibinfo {year}
  {2011})},\ \bibinfo {note} {particle and Nuclear Astrophysics International
  Workshop on Nuclear Physics, 32nd Course}\BibitemShut {NoStop}%
\bibitem [{\citenamefont {Cowan}\ \emph {et~al.}(1991)\citenamefont {Cowan},
  \citenamefont {Thielemann},\ and\ \citenamefont {Truran}}]{cowan91:_r}%
  \BibitemOpen
  \bibfield  {author} {\bibinfo {author} {\bibfnamefont {J.~J.}\ \bibnamefont
  {Cowan}}, \bibinfo {author} {\bibfnamefont {F.-K.}\ \bibnamefont
  {Thielemann}}, \ and\ \bibinfo {author} {\bibfnamefont {J.~W.}\ \bibnamefont
  {Truran}},\ }\href {\doibase 10.1016/0370-1573(91)90070-3} {\bibfield
  {journal} {\bibinfo  {journal} {Physics Reports}\ }\textbf {\bibinfo {volume}
  {208}},\ \bibinfo {pages} {267 } (\bibinfo {year} {1991})}\BibitemShut
  {NoStop}%
\bibitem [{\citenamefont {Schatz}(2013)}]{schatz13:_nuclear}%
  \BibitemOpen
  \bibfield  {author} {\bibinfo {author} {\bibfnamefont {H.}~\bibnamefont
  {Schatz}},\ }\href {\doibase 10.1016/j.ijms.2013.03.016} {\bibfield
  {journal} {\bibinfo  {journal} {International Journal of Mass Spectrometry}\
  }\textbf {\bibinfo {volume} {349-350}},\ \bibinfo {pages} {181 } (\bibinfo
  {year} {2013})},\ \bibinfo {note} {100 years of Mass
  Spectrometry}\BibitemShut {NoStop}%
\bibitem [{\citenamefont {Burbidge}\ \emph {et~al.}(1957)\citenamefont
  {Burbidge}, \citenamefont {Burbidge}, \citenamefont {Fowler},\ and\
  \citenamefont {Hoyle}}]{burbidge57:_synth_elemen_stars}%
  \BibitemOpen
  \bibfield  {author} {\bibinfo {author} {\bibfnamefont {E.~M.}\ \bibnamefont
  {Burbidge}}, \bibinfo {author} {\bibfnamefont {G.~R.}\ \bibnamefont
  {Burbidge}}, \bibinfo {author} {\bibfnamefont {W.~A.}\ \bibnamefont
  {Fowler}}, \ and\ \bibinfo {author} {\bibfnamefont {F.}~\bibnamefont
  {Hoyle}},\ }\href {\doibase 10.1103/RevModPhys.29.547} {\bibfield  {journal}
  {\bibinfo  {journal} {Rev. Mod. Phys.}\ }\textbf {\bibinfo {volume} {29}},\
  \bibinfo {pages} {547} (\bibinfo {year} {1957})}\BibitemShut {NoStop}%
\bibitem [{\citenamefont {Brett}\ \emph {et~al.}(2012)\citenamefont {Brett},
  \citenamefont {Bentley}, \citenamefont {Paul}, \citenamefont {Surman},\ and\
  \citenamefont {Aprahamian}}]{brett12:_sensit}%
  \BibitemOpen
  \bibfield  {author} {\bibinfo {author} {\bibfnamefont {S.}~\bibnamefont
  {Brett}}, \bibinfo {author} {\bibfnamefont {I.}~\bibnamefont {Bentley}},
  \bibinfo {author} {\bibfnamefont {N.}~\bibnamefont {Paul}}, \bibinfo {author}
  {\bibfnamefont {R.}~\bibnamefont {Surman}}, \ and\ \bibinfo {author}
  {\bibfnamefont {A.}~\bibnamefont {Aprahamian}},\ }\href {\doibase
  10.1140/epja/i2012-12184-4} {\bibfield  {journal} {\bibinfo  {journal} {The
  European Physical Journal A}\ }\textbf {\bibinfo {volume} {48}},\ \bibinfo
  {eid} {184} (\bibinfo {year} {2012}),\
  10.1140/epja/i2012-12184-4}\BibitemShut {NoStop}%
\bibitem [{\citenamefont {Moller}\ \emph {et~al.}(1995)\citenamefont {Moller},
  \citenamefont {Nix}, \citenamefont {Myers},\ and\ \citenamefont
  {Swiatecki}}]{moller95:_nuclear_groun_state_masses_defor}%
  \BibitemOpen
  \bibfield  {author} {\bibinfo {author} {\bibfnamefont {P.}~\bibnamefont
  {Moller}}, \bibinfo {author} {\bibfnamefont {J.}~\bibnamefont {Nix}},
  \bibinfo {author} {\bibfnamefont {W.}~\bibnamefont {Myers}}, \ and\ \bibinfo
  {author} {\bibfnamefont {W.}~\bibnamefont {Swiatecki}},\ }\href {\doibase
  10.1006/adnd.1995.1002} {\bibfield  {journal} {\bibinfo  {journal} {Atomic
  Data and Nuclear Data Tables}\ }\textbf {\bibinfo {volume} {59}},\ \bibinfo
  {pages} {185 } (\bibinfo {year} {1995})}\BibitemShut {NoStop}%
\bibitem [{\citenamefont {Duflo}\ and\ \citenamefont
  {Zuker}(1995)}]{duflo95:_micros}%
  \BibitemOpen
  \bibfield  {author} {\bibinfo {author} {\bibfnamefont {J.}~\bibnamefont
  {Duflo}}\ and\ \bibinfo {author} {\bibfnamefont {A.}~\bibnamefont {Zuker}},\
  }\href {\doibase 10.1103/PhysRevC.52.R23, 10.1103/PhysRevC.52.23} {\bibfield
  {journal} {\bibinfo  {journal} {Phys.Rev.}\ }\textbf {\bibinfo {volume}
  {C52}},\ \bibinfo {pages} {23} (\bibinfo {year} {1995})},\ \Eprint
  {http://arxiv.org/abs/nucl-th/9404019} {arXiv:nucl-th/9404019 [nucl-th]}
  \BibitemShut {NoStop}%
\bibitem [{\citenamefont {Pearson}\ \emph {et~al.}(1996)\citenamefont
  {Pearson}, \citenamefont {Nayak},\ and\ \citenamefont
  {Goriely}}]{pearson96:_nuclear_bogol}%
  \BibitemOpen
  \bibfield  {author} {\bibinfo {author} {\bibfnamefont {J.}~\bibnamefont
  {Pearson}}, \bibinfo {author} {\bibfnamefont {R.}~\bibnamefont {Nayak}}, \
  and\ \bibinfo {author} {\bibfnamefont {S.}~\bibnamefont {Goriely}},\ }\href
  {\doibase 10.1016/0370-2693(96)01071-4} {\bibfield  {journal} {\bibinfo
  {journal} {Physics Letters B}\ }\textbf {\bibinfo {volume} {387}},\ \bibinfo
  {pages} {455 } (\bibinfo {year} {1996})}\BibitemShut {NoStop}%
\bibitem [{\citenamefont {Rodriguez-Guzman}\ \emph {et~al.}(2010)\citenamefont
  {Rodriguez-Guzman}, \citenamefont {Sarriguren},\ and\ \citenamefont
  {Robledo}}]{rodriguez-guzman10:_signat}%
  \BibitemOpen
  \bibfield  {author} {\bibinfo {author} {\bibfnamefont {R.}~\bibnamefont
  {Rodriguez-Guzman}}, \bibinfo {author} {\bibfnamefont {P.}~\bibnamefont
  {Sarriguren}}, \ and\ \bibinfo {author} {\bibfnamefont {L.~M.}\ \bibnamefont
  {Robledo}},\ }\href {\doibase 10.1103/PhysRevC.82.061302} {\bibfield
  {journal} {\bibinfo  {journal} {Phys. Rev. C}\ }\textbf {\bibinfo {volume}
  {82}},\ \bibinfo {pages} {061302} (\bibinfo {year} {2010})}\BibitemShut
  {NoStop}%
\bibitem [{\citenamefont {Simon}\ \emph {et~al.}(2012)\citenamefont {Simon},
  \citenamefont {Brunner}, \citenamefont {Chowdhury}, \citenamefont
  {Eberhardt}, \citenamefont {Ettenauer}, \citenamefont {Gallant},
  \citenamefont {Man\'e}, \citenamefont {Simon}, \citenamefont {Delheij},
  \citenamefont {Pearson}, \citenamefont {Audi}, \citenamefont {Gwinner},
  \citenamefont {Lunney}, \citenamefont {Schatz},\ and\ \citenamefont
  {Dilling}}]{simon12:_pennin_rb_sr}%
  \BibitemOpen
  \bibfield  {author} {\bibinfo {author} {\bibfnamefont {V.~V.}\ \bibnamefont
  {Simon}}, \bibinfo {author} {\bibfnamefont {T.}~\bibnamefont {Brunner}},
  \bibinfo {author} {\bibfnamefont {U.}~\bibnamefont {Chowdhury}}, \bibinfo
  {author} {\bibfnamefont {B.}~\bibnamefont {Eberhardt}}, \bibinfo {author}
  {\bibfnamefont {S.}~\bibnamefont {Ettenauer}}, \bibinfo {author}
  {\bibfnamefont {A.~T.}\ \bibnamefont {Gallant}}, \bibinfo {author}
  {\bibfnamefont {E.}~\bibnamefont {Man\'e}}, \bibinfo {author} {\bibfnamefont
  {M.~C.}\ \bibnamefont {Simon}}, \bibinfo {author} {\bibfnamefont
  {P.}~\bibnamefont {Delheij}}, \bibinfo {author} {\bibfnamefont {M.~R.}\
  \bibnamefont {Pearson}}, \bibinfo {author} {\bibfnamefont {G.}~\bibnamefont
  {Audi}}, \bibinfo {author} {\bibfnamefont {G.}~\bibnamefont {Gwinner}},
  \bibinfo {author} {\bibfnamefont {D.}~\bibnamefont {Lunney}}, \bibinfo
  {author} {\bibfnamefont {H.}~\bibnamefont {Schatz}}, \ and\ \bibinfo {author}
  {\bibfnamefont {J.}~\bibnamefont {Dilling}},\ }\href {\doibase
  10.1103/PhysRevC.85.064308} {\bibfield  {journal} {\bibinfo  {journal} {Phys.
  Rev. C}\ }\textbf {\bibinfo {volume} {85}},\ \bibinfo {pages} {064308}
  (\bibinfo {year} {2012})}\BibitemShut {NoStop}%
\bibitem [{\citenamefont {Blaum}\ \emph {et~al.}(2013)\citenamefont {Blaum},
  \citenamefont {Dilling},\ and\ \citenamefont
  {N{\"o}rtershauser}}]{blaum13:_precis_atomic_physic_techn_nuclear}%
  \BibitemOpen
  \bibfield  {author} {\bibinfo {author} {\bibfnamefont {K.}~\bibnamefont
  {Blaum}}, \bibinfo {author} {\bibfnamefont {J.}~\bibnamefont {Dilling}}, \
  and\ \bibinfo {author} {\bibfnamefont {W.}~\bibnamefont
  {N{\"o}rtershauser}},\ }\href {\doibase 10.1088/0031-8949/2013/T152/014017}
  {\bibfield  {journal} {\bibinfo  {journal} {Phys.Scripta}\ }\textbf {\bibinfo
  {volume} {T152}},\ \bibinfo {pages} {014017} (\bibinfo {year} {2013})},\
  \Eprint {http://arxiv.org/abs/1210.4045} {arXiv:1210.4045 [physics.atom-ph]}
  \BibitemShut {NoStop}%
\bibitem [{\citenamefont {Kwiatkowski}\ \emph {et~al.}(2014)\citenamefont
  {Kwiatkowski}, \citenamefont {Andreoiu}, \citenamefont {Bale}, \citenamefont
  {Brunner}, \citenamefont {Chaudhuri}, \citenamefont {Chowdhury},
  \citenamefont {Delheij}, \citenamefont {Ettenauer}, \citenamefont {Frekers},
  \citenamefont {Gallant}, \citenamefont {Grossheim}, \citenamefont {Gwinner},
  \citenamefont {Jang}, \citenamefont {Lennarz}, \citenamefont {Ma},
  \citenamefont {Man{\'e}}, \citenamefont {Pearson}, \citenamefont {Schultz},
  \citenamefont {Simon}, \citenamefont {Simon},\ and\ \citenamefont
  {Dilling}}]{kwiatkowski14:_titan}%
  \BibitemOpen
  \bibfield  {author} {\bibinfo {author} {\bibfnamefont {A.~A.}\ \bibnamefont
  {Kwiatkowski}}, \bibinfo {author} {\bibfnamefont {C.}~\bibnamefont
  {Andreoiu}}, \bibinfo {author} {\bibfnamefont {J.~C.}\ \bibnamefont {Bale}},
  \bibinfo {author} {\bibfnamefont {T.}~\bibnamefont {Brunner}}, \bibinfo
  {author} {\bibfnamefont {A.}~\bibnamefont {Chaudhuri}}, \bibinfo {author}
  {\bibfnamefont {U.}~\bibnamefont {Chowdhury}}, \bibinfo {author}
  {\bibfnamefont {P.}~\bibnamefont {Delheij}}, \bibinfo {author} {\bibfnamefont
  {S.}~\bibnamefont {Ettenauer}}, \bibinfo {author} {\bibfnamefont
  {D.}~\bibnamefont {Frekers}}, \bibinfo {author} {\bibfnamefont {A.~T.}\
  \bibnamefont {Gallant}}, \bibinfo {author} {\bibfnamefont {A.}~\bibnamefont
  {Grossheim}}, \bibinfo {author} {\bibfnamefont {G.}~\bibnamefont {Gwinner}},
  \bibinfo {author} {\bibfnamefont {F.}~\bibnamefont {Jang}}, \bibinfo {author}
  {\bibfnamefont {A.}~\bibnamefont {Lennarz}}, \bibinfo {author} {\bibfnamefont
  {T.}~\bibnamefont {Ma}}, \bibinfo {author} {\bibfnamefont {E.}~\bibnamefont
  {Man{\'e}}}, \bibinfo {author} {\bibfnamefont {M.~R.}\ \bibnamefont
  {Pearson}}, \bibinfo {author} {\bibfnamefont {B.~E.}\ \bibnamefont
  {Schultz}}, \bibinfo {author} {\bibfnamefont {M.~C.}\ \bibnamefont {Simon}},
  \bibinfo {author} {\bibfnamefont {V.~V.}\ \bibnamefont {Simon}}, \ and\
  \bibinfo {author} {\bibfnamefont {J.}~\bibnamefont {Dilling}},\ }\href
  {\doibase 10.1007/s10751-013-0892-8} {\bibfield  {journal} {\bibinfo
  {journal} {Hyperfine Interactions}\ }\textbf {\bibinfo {volume} {225}},\
  \bibinfo {pages} {143} (\bibinfo {year} {2014})}\BibitemShut {NoStop}%
\bibitem [{\citenamefont {Dilling}\ \emph {et~al.}(2006)\citenamefont
  {Dilling}, \citenamefont {Baartman}, \citenamefont {Bricault}, \citenamefont
  {Brodeur}, \citenamefont {Blomeley}, \citenamefont {Buchinger}, \citenamefont
  {Crawford}, \citenamefont {L{\'o}pez-Urrutia}, \citenamefont {Delheij},
  \citenamefont {Froese}, \citenamefont {Gwinner}, \citenamefont {Ke},
  \citenamefont {Lee}, \citenamefont {Moore}, \citenamefont {Ryjkov},
  \citenamefont {Sikler}, \citenamefont {Smith}, \citenamefont {Ullrich},\ and\
  \citenamefont {Vaz}}]{dilling06:_mass_titan}%
  \BibitemOpen
  \bibfield  {author} {\bibinfo {author} {\bibfnamefont {J.}~\bibnamefont
  {Dilling}}, \bibinfo {author} {\bibfnamefont {R.}~\bibnamefont {Baartman}},
  \bibinfo {author} {\bibfnamefont {P.}~\bibnamefont {Bricault}}, \bibinfo
  {author} {\bibfnamefont {M.}~\bibnamefont {Brodeur}}, \bibinfo {author}
  {\bibfnamefont {L.}~\bibnamefont {Blomeley}}, \bibinfo {author}
  {\bibfnamefont {F.}~\bibnamefont {Buchinger}}, \bibinfo {author}
  {\bibfnamefont {J.}~\bibnamefont {Crawford}}, \bibinfo {author}
  {\bibfnamefont {J.~C.}\ \bibnamefont {L{\'o}pez-Urrutia}}, \bibinfo {author}
  {\bibfnamefont {P.}~\bibnamefont {Delheij}}, \bibinfo {author} {\bibfnamefont
  {M.}~\bibnamefont {Froese}}, \bibinfo {author} {\bibfnamefont
  {G.}~\bibnamefont {Gwinner}}, \bibinfo {author} {\bibfnamefont
  {Z.}~\bibnamefont {Ke}}, \bibinfo {author} {\bibfnamefont {J.}~\bibnamefont
  {Lee}}, \bibinfo {author} {\bibfnamefont {R.}~\bibnamefont {Moore}}, \bibinfo
  {author} {\bibfnamefont {V.}~\bibnamefont {Ryjkov}}, \bibinfo {author}
  {\bibfnamefont {G.}~\bibnamefont {Sikler}}, \bibinfo {author} {\bibfnamefont
  {M.}~\bibnamefont {Smith}}, \bibinfo {author} {\bibfnamefont
  {J.}~\bibnamefont {Ullrich}}, \ and\ \bibinfo {author} {\bibfnamefont
  {J.}~\bibnamefont {Vaz}},\ }\href {\doibase 10.1016/j.ijms.2006.01.044}
  {\bibfield  {journal} {\bibinfo  {journal} {International Journal of Mass
  Spectrometry}\ }\textbf {\bibinfo {volume} {251}},\ \bibinfo {pages} {198 }
  (\bibinfo {year} {2006})}\BibitemShut {NoStop}%
\bibitem [{\citenamefont {Dilling}\ \emph {et~al.}(2014)\citenamefont
  {Dilling}, \citenamefont {Kr{\"u}cken},\ and\ \citenamefont
  {Ball}}]{dilling14:_isac}%
  \BibitemOpen
  \bibfield  {author} {\bibinfo {author} {\bibfnamefont {J.}~\bibnamefont
  {Dilling}}, \bibinfo {author} {\bibfnamefont {R.}~\bibnamefont
  {Kr{\"u}cken}}, \ and\ \bibinfo {author} {\bibfnamefont {G.}~\bibnamefont
  {Ball}},\ }in\ \href {\doibase 10.1007/978-94-007-7963-1_1} {\emph {\bibinfo
  {booktitle} {ISAC and ARIEL: The TRIUMF Radioactive Beam Facilities and the
  Scientific Program}}},\ \bibinfo {editor} {edited by\ \bibinfo {editor}
  {\bibfnamefont {J.}~\bibnamefont {Dilling}}, \bibinfo {editor} {\bibfnamefont
  {R.}~\bibnamefont {Kr{\"u}cken}}, \ and\ \bibinfo {editor} {\bibfnamefont
  {L.}~\bibnamefont {Merminga}}}\ (\bibinfo  {publisher} {Springer
  Netherlands},\ \bibinfo {year} {2014})\ pp.\ \bibinfo {pages}
  {1--8}\BibitemShut {NoStop}%
\bibitem [{\citenamefont {Kunz}\ \emph {et~al.}(2013)\citenamefont {Kunz},
  \citenamefont {Bricault}, \citenamefont {Dombsky}, \citenamefont {Erdmann},
  \citenamefont {Hanemaayer}, \citenamefont {Wong},\ and\ \citenamefont
  {L{\"u}tzenkirchen}}]{kunz13:_compos_trium}%
  \BibitemOpen
  \bibfield  {author} {\bibinfo {author} {\bibfnamefont {P.}~\bibnamefont
  {Kunz}}, \bibinfo {author} {\bibfnamefont {P.}~\bibnamefont {Bricault}},
  \bibinfo {author} {\bibfnamefont {M.}~\bibnamefont {Dombsky}}, \bibinfo
  {author} {\bibfnamefont {N.}~\bibnamefont {Erdmann}}, \bibinfo {author}
  {\bibfnamefont {V.}~\bibnamefont {Hanemaayer}}, \bibinfo {author}
  {\bibfnamefont {J.}~\bibnamefont {Wong}}, \ and\ \bibinfo {author}
  {\bibfnamefont {K.}~\bibnamefont {L{\"u}tzenkirchen}},\ }\href {\doibase
  10.1016/j.jnucmat.2013.04.065} {\bibfield  {journal} {\bibinfo  {journal}
  {Journal of Nuclear Materials}\ }\textbf {\bibinfo {volume} {440}},\ \bibinfo
  {pages} {110 } (\bibinfo {year} {2013})}\BibitemShut {NoStop}%
\bibitem [{\citenamefont {Bricault}\ \emph {et~al.}(2014)\citenamefont
  {Bricault}, \citenamefont {Ames}, \citenamefont {Dombsky}, \citenamefont
  {Kunz},\ and\ \citenamefont {Lassen}}]{dilling14:_rare_isac}%
  \BibitemOpen
  \bibfield  {author} {\bibinfo {author} {\bibfnamefont {P.}~\bibnamefont
  {Bricault}}, \bibinfo {author} {\bibfnamefont {F.}~\bibnamefont {Ames}},
  \bibinfo {author} {\bibfnamefont {M.}~\bibnamefont {Dombsky}}, \bibinfo
  {author} {\bibfnamefont {P.}~\bibnamefont {Kunz}}, \ and\ \bibinfo {author}
  {\bibfnamefont {J.}~\bibnamefont {Lassen}},\ }in\ \href {\doibase
  10.1007/978-94-007-7963-1_4} {\emph {\bibinfo {booktitle} {ISAC and ARIEL:
  The TRIUMF Radioactive Beam Facilities and the Scientific Program}}}\
  (\bibinfo  {publisher} {Springer Netherlands},\ \bibinfo {year} {2014})\ pp.\
  \bibinfo {pages} {25--49}\BibitemShut {NoStop}%
\bibitem [{\citenamefont {Brunner}\ \emph
  {et~al.}(2012{\natexlab{a}})\citenamefont {Brunner}, \citenamefont {Smith},
  \citenamefont {Brodeur}, \citenamefont {Ettenauer}, \citenamefont {Gallant},
  \citenamefont {Simon}, \citenamefont {Chaudhuri}, \citenamefont {Lapierre},
  \citenamefont {Man{\'e}}, \citenamefont {Ringle}, \citenamefont {Simon},
  \citenamefont {Vaz}, \citenamefont {Delheij}, \citenamefont {Good},
  \citenamefont {Pearson},\ and\ \citenamefont
  {Dilling}}]{brunner12:_titan_rfq}%
  \BibitemOpen
  \bibfield  {author} {\bibinfo {author} {\bibfnamefont {T.}~\bibnamefont
  {Brunner}}, \bibinfo {author} {\bibfnamefont {M.}~\bibnamefont {Smith}},
  \bibinfo {author} {\bibfnamefont {M.}~\bibnamefont {Brodeur}}, \bibinfo
  {author} {\bibfnamefont {S.}~\bibnamefont {Ettenauer}}, \bibinfo {author}
  {\bibfnamefont {A.}~\bibnamefont {Gallant}}, \bibinfo {author} {\bibfnamefont
  {V.}~\bibnamefont {Simon}}, \bibinfo {author} {\bibfnamefont
  {A.}~\bibnamefont {Chaudhuri}}, \bibinfo {author} {\bibfnamefont
  {A.}~\bibnamefont {Lapierre}}, \bibinfo {author} {\bibfnamefont
  {E.}~\bibnamefont {Man{\'e}}}, \bibinfo {author} {\bibfnamefont
  {R.}~\bibnamefont {Ringle}}, \bibinfo {author} {\bibfnamefont
  {M.}~\bibnamefont {Simon}}, \bibinfo {author} {\bibfnamefont
  {J.}~\bibnamefont {Vaz}}, \bibinfo {author} {\bibfnamefont {P.}~\bibnamefont
  {Delheij}}, \bibinfo {author} {\bibfnamefont {M.}~\bibnamefont {Good}},
  \bibinfo {author} {\bibfnamefont {M.}~\bibnamefont {Pearson}}, \ and\
  \bibinfo {author} {\bibfnamefont {J.}~\bibnamefont {Dilling}},\ }\href
  {\doibase 10.1016/j.nima.2012.02.004} {\bibfield  {journal} {\bibinfo
  {journal} {Nuclear Instruments and Methods in Physics Research Section A:
  Accelerators, Spectrometers, Detectors and Associated Equipment}\ }\textbf
  {\bibinfo {volume} {676}},\ \bibinfo {pages} {32 } (\bibinfo {year}
  {2012}{\natexlab{a}})}\BibitemShut {NoStop}%
\bibitem [{\citenamefont {Smith}\ \emph {et~al.}(2006)\citenamefont {Smith},
  \citenamefont {Blomeley}, \citenamefont {Delheij},\ and\ \citenamefont
  {Dilling}}]{smith06:_first_titan_rfq}%
  \BibitemOpen
  \bibfield  {author} {\bibinfo {author} {\bibfnamefont {M.}~\bibnamefont
  {Smith}}, \bibinfo {author} {\bibfnamefont {L.}~\bibnamefont {Blomeley}},
  \bibinfo {author} {\bibfnamefont {P.}~\bibnamefont {Delheij}}, \ and\
  \bibinfo {author} {\bibfnamefont {J.}~\bibnamefont {Dilling}},\ }\href
  {\doibase 10.1007/s10751-007-9554-z} {\bibfield  {journal} {\bibinfo
  {journal} {Hyperfine Interactions}\ }\textbf {\bibinfo {volume} {173}},\
  \bibinfo {pages} {171} (\bibinfo {year} {2006})}\BibitemShut {NoStop}%
\bibitem [{\citenamefont {Lapierre}\ \emph {et~al.}(2010)\citenamefont
  {Lapierre}, \citenamefont {Brodeur}, \citenamefont {Brunner}, \citenamefont
  {Ettenauer}, \citenamefont {Gallant}, \citenamefont {Simon}, \citenamefont
  {Good}, \citenamefont {Froese}, \citenamefont {L{\'o}pez-Urrutia},
  \citenamefont {Delheij}, \citenamefont {Epp}, \citenamefont {Ringle},
  \citenamefont {Schwarz}, \citenamefont {Ullrich},\ and\ \citenamefont
  {Dilling}}]{lapierre10:_titan_ebit}%
  \BibitemOpen
  \bibfield  {author} {\bibinfo {author} {\bibfnamefont {A.}~\bibnamefont
  {Lapierre}}, \bibinfo {author} {\bibfnamefont {M.}~\bibnamefont {Brodeur}},
  \bibinfo {author} {\bibfnamefont {T.}~\bibnamefont {Brunner}}, \bibinfo
  {author} {\bibfnamefont {S.}~\bibnamefont {Ettenauer}}, \bibinfo {author}
  {\bibfnamefont {A.}~\bibnamefont {Gallant}}, \bibinfo {author} {\bibfnamefont
  {V.}~\bibnamefont {Simon}}, \bibinfo {author} {\bibfnamefont
  {M.}~\bibnamefont {Good}}, \bibinfo {author} {\bibfnamefont {M.}~\bibnamefont
  {Froese}}, \bibinfo {author} {\bibfnamefont {J.~C.}\ \bibnamefont
  {L{\'o}pez-Urrutia}}, \bibinfo {author} {\bibfnamefont {P.}~\bibnamefont
  {Delheij}}, \bibinfo {author} {\bibfnamefont {S.}~\bibnamefont {Epp}},
  \bibinfo {author} {\bibfnamefont {R.}~\bibnamefont {Ringle}}, \bibinfo
  {author} {\bibfnamefont {S.}~\bibnamefont {Schwarz}}, \bibinfo {author}
  {\bibfnamefont {J.}~\bibnamefont {Ullrich}}, \ and\ \bibinfo {author}
  {\bibfnamefont {J.}~\bibnamefont {Dilling}},\ }\href {\doibase
  10.1016/j.nima.2010.09.030} {\bibfield  {journal} {\bibinfo  {journal}
  {Nuclear Instruments and Methods in Physics Research Section A: Accelerators,
  Spectrometers, Detectors and Associated Equipment}\ }\textbf {\bibinfo
  {volume} {624}},\ \bibinfo {pages} {54 } (\bibinfo {year}
  {2010})}\BibitemShut {NoStop}%
\bibitem [{\citenamefont {Brunner}\ \emph
  {et~al.}(2012{\natexlab{b}})\citenamefont {Brunner}, \citenamefont {Mueller},
  \citenamefont {O'Sullivan}, \citenamefont {Simon}, \citenamefont {Kossick},
  \citenamefont {Ettenauer}, \citenamefont {Gallant}, \citenamefont {Man{\'e}},
  \citenamefont {Bishop}, \citenamefont {Good}, \citenamefont {Gratta},\ and\
  \citenamefont {Dilling}}]{brunner12:_bradb_niels}%
  \BibitemOpen
  \bibfield  {author} {\bibinfo {author} {\bibfnamefont {T.}~\bibnamefont
  {Brunner}}, \bibinfo {author} {\bibfnamefont {A.}~\bibnamefont {Mueller}},
  \bibinfo {author} {\bibfnamefont {K.}~\bibnamefont {O'Sullivan}}, \bibinfo
  {author} {\bibfnamefont {M.}~\bibnamefont {Simon}}, \bibinfo {author}
  {\bibfnamefont {M.}~\bibnamefont {Kossick}}, \bibinfo {author} {\bibfnamefont
  {S.}~\bibnamefont {Ettenauer}}, \bibinfo {author} {\bibfnamefont
  {A.}~\bibnamefont {Gallant}}, \bibinfo {author} {\bibfnamefont
  {E.}~\bibnamefont {Man{\'e}}}, \bibinfo {author} {\bibfnamefont
  {D.}~\bibnamefont {Bishop}}, \bibinfo {author} {\bibfnamefont
  {M.}~\bibnamefont {Good}}, \bibinfo {author} {\bibfnamefont {G.}~\bibnamefont
  {Gratta}}, \ and\ \bibinfo {author} {\bibfnamefont {J.}~\bibnamefont
  {Dilling}},\ }\href {\doibase 10.1016/j.ijms.2011.09.004} {\bibfield
  {journal} {\bibinfo  {journal} {International Journal of Mass Spectrometry}\
  }\textbf {\bibinfo {volume} {309}},\ \bibinfo {pages} {97 } (\bibinfo {year}
  {2012}{\natexlab{b}})}\BibitemShut {NoStop}%
\bibitem [{\citenamefont {K{\"o}nig}\ \emph {et~al.}(1995)\citenamefont
  {K{\"o}nig}, \citenamefont {Bollen}, \citenamefont {Kluge}, \citenamefont
  {Otto},\ and\ \citenamefont {Szerypo}}]{koenig95:_quadr}%
  \BibitemOpen
  \bibfield  {author} {\bibinfo {author} {\bibfnamefont {M.}~\bibnamefont
  {K{\"o}nig}}, \bibinfo {author} {\bibfnamefont {G.}~\bibnamefont {Bollen}},
  \bibinfo {author} {\bibfnamefont {H.-J.}\ \bibnamefont {Kluge}}, \bibinfo
  {author} {\bibfnamefont {T.}~\bibnamefont {Otto}}, \ and\ \bibinfo {author}
  {\bibfnamefont {J.}~\bibnamefont {Szerypo}},\ }\href {\doibase
  10.1016/0168-1176(95)04146-C} {\bibfield  {journal} {\bibinfo  {journal}
  {International Journal of Mass Spectrometry and Ion Processes}\ }\textbf
  {\bibinfo {volume} {142}},\ \bibinfo {pages} {95 } (\bibinfo {year}
  {1995})}\BibitemShut {NoStop}%
\bibitem [{\citenamefont {Brodeur}\ \emph {et~al.}(2012)\citenamefont
  {Brodeur}, \citenamefont {Ryjkov}, \citenamefont {Brunner}, \citenamefont
  {Ettenauer}, \citenamefont {Gallant}, \citenamefont {Simon}, \citenamefont
  {Smith}, \citenamefont {Lapierre}, \citenamefont {Ringle}, \citenamefont
  {Delheij}, \citenamefont {Good}, \citenamefont {Lunney},\ and\ \citenamefont
  {Dilling}}]{brodeur12:_verif_titan_pennin}%
  \BibitemOpen
  \bibfield  {author} {\bibinfo {author} {\bibfnamefont {M.}~\bibnamefont
  {Brodeur}}, \bibinfo {author} {\bibfnamefont {V.}~\bibnamefont {Ryjkov}},
  \bibinfo {author} {\bibfnamefont {T.}~\bibnamefont {Brunner}}, \bibinfo
  {author} {\bibfnamefont {S.}~\bibnamefont {Ettenauer}}, \bibinfo {author}
  {\bibfnamefont {A.}~\bibnamefont {Gallant}}, \bibinfo {author} {\bibfnamefont
  {V.}~\bibnamefont {Simon}}, \bibinfo {author} {\bibfnamefont
  {M.}~\bibnamefont {Smith}}, \bibinfo {author} {\bibfnamefont
  {A.}~\bibnamefont {Lapierre}}, \bibinfo {author} {\bibfnamefont
  {R.}~\bibnamefont {Ringle}}, \bibinfo {author} {\bibfnamefont
  {P.}~\bibnamefont {Delheij}}, \bibinfo {author} {\bibfnamefont
  {M.}~\bibnamefont {Good}}, \bibinfo {author} {\bibfnamefont {D.}~\bibnamefont
  {Lunney}}, \ and\ \bibinfo {author} {\bibfnamefont {J.}~\bibnamefont
  {Dilling}},\ }\href {\doibase 10.1016/j.ijms.2011.11.002} {\bibfield
  {journal} {\bibinfo  {journal} {International Journal of Mass Spectrometry}\
  }\textbf {\bibinfo {volume} {310}},\ \bibinfo {pages} {20 } (\bibinfo {year}
  {2012})}\BibitemShut {NoStop}%
\bibitem [{\citenamefont {Ettenauer}\ \emph {et~al.}(2011)\citenamefont
  {Ettenauer}, \citenamefont {Simon}, \citenamefont {Gallant}, \citenamefont
  {Brunner}, \citenamefont {Chowdhury}, \citenamefont {Simon}, \citenamefont
  {Brodeur}, \citenamefont {Chaudhuri}, \citenamefont {Man\'e}, \citenamefont
  {Andreoiu}, \citenamefont {Audi}, \citenamefont {L\'opez-Urrutia},
  \citenamefont {Delheij}, \citenamefont {Gwinner}, \citenamefont {Lapierre},
  \citenamefont {Lunney}, \citenamefont {Pearson}, \citenamefont {Ringle},
  \citenamefont {Ullrich},\ and\ \citenamefont
  {Dilling}}]{ettenauer11:_first_use_high_charg_states}%
  \BibitemOpen
  \bibfield  {author} {\bibinfo {author} {\bibfnamefont {S.}~\bibnamefont
  {Ettenauer}}, \bibinfo {author} {\bibfnamefont {M.~C.}\ \bibnamefont
  {Simon}}, \bibinfo {author} {\bibfnamefont {A.~T.}\ \bibnamefont {Gallant}},
  \bibinfo {author} {\bibfnamefont {T.}~\bibnamefont {Brunner}}, \bibinfo
  {author} {\bibfnamefont {U.}~\bibnamefont {Chowdhury}}, \bibinfo {author}
  {\bibfnamefont {V.~V.}\ \bibnamefont {Simon}}, \bibinfo {author}
  {\bibfnamefont {M.}~\bibnamefont {Brodeur}}, \bibinfo {author} {\bibfnamefont
  {A.}~\bibnamefont {Chaudhuri}}, \bibinfo {author} {\bibfnamefont
  {E.}~\bibnamefont {Man\'e}}, \bibinfo {author} {\bibfnamefont
  {C.}~\bibnamefont {Andreoiu}}, \bibinfo {author} {\bibfnamefont
  {G.}~\bibnamefont {Audi}}, \bibinfo {author} {\bibfnamefont {J.~R.~C.}\
  \bibnamefont {L\'opez-Urrutia}}, \bibinfo {author} {\bibfnamefont
  {P.}~\bibnamefont {Delheij}}, \bibinfo {author} {\bibfnamefont
  {G.}~\bibnamefont {Gwinner}}, \bibinfo {author} {\bibfnamefont
  {A.}~\bibnamefont {Lapierre}}, \bibinfo {author} {\bibfnamefont
  {D.}~\bibnamefont {Lunney}}, \bibinfo {author} {\bibfnamefont {M.~R.}\
  \bibnamefont {Pearson}}, \bibinfo {author} {\bibfnamefont {R.}~\bibnamefont
  {Ringle}}, \bibinfo {author} {\bibfnamefont {J.}~\bibnamefont {Ullrich}}, \
  and\ \bibinfo {author} {\bibfnamefont {J.}~\bibnamefont {Dilling}},\ }\href
  {\doibase 10.1103/PhysRevLett.107.272501} {\bibfield  {journal} {\bibinfo
  {journal} {Phys. Rev. Lett.}\ }\textbf {\bibinfo {volume} {107}},\ \bibinfo
  {pages} {272501} (\bibinfo {year} {2011})}\BibitemShut {NoStop}%
\bibitem [{\citenamefont {Malbrunot-Ettenauer}\ \emph
  {et~al.}(2015)\citenamefont {Malbrunot-Ettenauer}, \citenamefont {Brunner},
  \citenamefont {Chowdhury}, \citenamefont {Gallant}, \citenamefont {Simon},
  \citenamefont {Brodeur}, \citenamefont {Chaudhuri}, \citenamefont {Man\'e},
  \citenamefont {Simon}, \citenamefont {Andreoiu}, \citenamefont {Audi},
  \citenamefont {Crespo L\'opez-Urrutia}, \citenamefont {Delheij},
  \citenamefont {Gwinner}, \citenamefont {Lapierre}, \citenamefont {Lunney},
  \citenamefont {Pearson}, \citenamefont {Ringle}, \citenamefont {Ullrich},\
  and\ \citenamefont {Dilling}}]{malbrunot-ettenauer15:_pennin}%
  \BibitemOpen
  \bibfield  {author} {\bibinfo {author} {\bibfnamefont {S.}~\bibnamefont
  {Malbrunot-Ettenauer}}, \bibinfo {author} {\bibfnamefont {T.}~\bibnamefont
  {Brunner}}, \bibinfo {author} {\bibfnamefont {U.}~\bibnamefont {Chowdhury}},
  \bibinfo {author} {\bibfnamefont {A.~T.}\ \bibnamefont {Gallant}}, \bibinfo
  {author} {\bibfnamefont {V.~V.}\ \bibnamefont {Simon}}, \bibinfo {author}
  {\bibfnamefont {M.}~\bibnamefont {Brodeur}}, \bibinfo {author} {\bibfnamefont
  {A.}~\bibnamefont {Chaudhuri}}, \bibinfo {author} {\bibfnamefont
  {E.}~\bibnamefont {Man\'e}}, \bibinfo {author} {\bibfnamefont {M.~C.}\
  \bibnamefont {Simon}}, \bibinfo {author} {\bibfnamefont {C.}~\bibnamefont
  {Andreoiu}}, \bibinfo {author} {\bibfnamefont {G.}~\bibnamefont {Audi}},
  \bibinfo {author} {\bibfnamefont {J.~R.}\ \bibnamefont {Crespo
  L\'opez-Urrutia}}, \bibinfo {author} {\bibfnamefont {P.}~\bibnamefont
  {Delheij}}, \bibinfo {author} {\bibfnamefont {G.}~\bibnamefont {Gwinner}},
  \bibinfo {author} {\bibfnamefont {A.}~\bibnamefont {Lapierre}}, \bibinfo
  {author} {\bibfnamefont {D.}~\bibnamefont {Lunney}}, \bibinfo {author}
  {\bibfnamefont {M.~R.}\ \bibnamefont {Pearson}}, \bibinfo {author}
  {\bibfnamefont {R.}~\bibnamefont {Ringle}}, \bibinfo {author} {\bibfnamefont
  {J.}~\bibnamefont {Ullrich}}, \ and\ \bibinfo {author} {\bibfnamefont
  {J.}~\bibnamefont {Dilling}},\ }\href {\doibase 10.1103/PhysRevC.91.045504}
  {\bibfield  {journal} {\bibinfo  {journal} {Phys. Rev. C}\ }\textbf {\bibinfo
  {volume} {91}},\ \bibinfo {pages} {045504} (\bibinfo {year}
  {2015})}\BibitemShut {NoStop}%
\bibitem [{\citenamefont {Frekers}\ \emph {et~al.}(2013)\citenamefont
  {Frekers}, \citenamefont {Simon}, \citenamefont {Andreoiu}, \citenamefont
  {Bale}, \citenamefont {Brodeur}, \citenamefont {Brunner}, \citenamefont
  {Chaudhuri}, \citenamefont {Chowdhury}, \citenamefont {L{\'o}pez-Urrutia},
  \citenamefont {Delheij}, \citenamefont {Ejiri}, \citenamefont {Ettenauer},
  \citenamefont {Gallant}, \citenamefont {Gavrin}, \citenamefont {Grossheim},
  \citenamefont {Harakeh}, \citenamefont {Jang}, \citenamefont {Kwiatkowski},
  \citenamefont {Lassen}, \citenamefont {Lennarz}, \citenamefont {Luichtl},
  \citenamefont {Ma}, \citenamefont {Macdonald}, \citenamefont {Man{\'e}},
  \citenamefont {Robertson}, \citenamefont {Schultz}, \citenamefont {Simon},
  \citenamefont {Teigelh{\"o}fer},\ and\ \citenamefont
  {Dilling}}]{frekers13:_pennin_q}%
  \BibitemOpen
  \bibfield  {author} {\bibinfo {author} {\bibfnamefont {D.}~\bibnamefont
  {Frekers}}, \bibinfo {author} {\bibfnamefont {M.}~\bibnamefont {Simon}},
  \bibinfo {author} {\bibfnamefont {C.}~\bibnamefont {Andreoiu}}, \bibinfo
  {author} {\bibfnamefont {J.~C.}\ \bibnamefont {Bale}}, \bibinfo {author}
  {\bibfnamefont {M.}~\bibnamefont {Brodeur}}, \bibinfo {author} {\bibfnamefont
  {T.}~\bibnamefont {Brunner}}, \bibinfo {author} {\bibfnamefont
  {A.}~\bibnamefont {Chaudhuri}}, \bibinfo {author} {\bibfnamefont
  {U.}~\bibnamefont {Chowdhury}}, \bibinfo {author} {\bibfnamefont {J.~C.}\
  \bibnamefont {L{\'o}pez-Urrutia}}, \bibinfo {author} {\bibfnamefont
  {P.}~\bibnamefont {Delheij}}, \bibinfo {author} {\bibfnamefont
  {H.}~\bibnamefont {Ejiri}}, \bibinfo {author} {\bibfnamefont
  {S.}~\bibnamefont {Ettenauer}}, \bibinfo {author} {\bibfnamefont {A.~T.}\
  \bibnamefont {Gallant}}, \bibinfo {author} {\bibfnamefont {V.}~\bibnamefont
  {Gavrin}}, \bibinfo {author} {\bibfnamefont {A.}~\bibnamefont {Grossheim}},
  \bibinfo {author} {\bibfnamefont {M.~N.}\ \bibnamefont {Harakeh}}, \bibinfo
  {author} {\bibfnamefont {F.}~\bibnamefont {Jang}}, \bibinfo {author}
  {\bibfnamefont {A.~A.}\ \bibnamefont {Kwiatkowski}}, \bibinfo {author}
  {\bibfnamefont {J.}~\bibnamefont {Lassen}}, \bibinfo {author} {\bibfnamefont
  {A.}~\bibnamefont {Lennarz}}, \bibinfo {author} {\bibfnamefont
  {M.}~\bibnamefont {Luichtl}}, \bibinfo {author} {\bibfnamefont
  {T.}~\bibnamefont {Ma}}, \bibinfo {author} {\bibfnamefont {T.~D.}\
  \bibnamefont {Macdonald}}, \bibinfo {author} {\bibfnamefont {E.}~\bibnamefont
  {Man{\'e}}}, \bibinfo {author} {\bibfnamefont {D.}~\bibnamefont {Robertson}},
  \bibinfo {author} {\bibfnamefont {B.}~\bibnamefont {Schultz}}, \bibinfo
  {author} {\bibfnamefont {V.~V.}\ \bibnamefont {Simon}}, \bibinfo {author}
  {\bibfnamefont {A.}~\bibnamefont {Teigelh{\"o}fer}}, \ and\ \bibinfo {author}
  {\bibfnamefont {J.}~\bibnamefont {Dilling}},\ }\href {\doibase
  10.1016/j.physletb.2013.04.019} {\bibfield  {journal} {\bibinfo  {journal}
  {Physics Letters B}\ }\textbf {\bibinfo {volume} {722}},\ \bibinfo {pages}
  {233 } (\bibinfo {year} {2013})}\BibitemShut {NoStop}%
\bibitem [{\citenamefont {Ettenauer}\ \emph {et~al.}(2013)\citenamefont
  {Ettenauer}, \citenamefont {Simon}, \citenamefont {Macdonald},\ and\
  \citenamefont {Dilling}}]{ettenauer13:_advan_pennin}%
  \BibitemOpen
  \bibfield  {author} {\bibinfo {author} {\bibfnamefont {S.}~\bibnamefont
  {Ettenauer}}, \bibinfo {author} {\bibfnamefont {M.}~\bibnamefont {Simon}},
  \bibinfo {author} {\bibfnamefont {T.}~\bibnamefont {Macdonald}}, \ and\
  \bibinfo {author} {\bibfnamefont {J.}~\bibnamefont {Dilling}},\ }\href
  {\doibase 10.1016/j.ijms.2013.04.021} {\bibfield  {journal} {\bibinfo
  {journal} {International Journal of Mass Spectrometry}\ }\textbf {\bibinfo
  {volume} {349 - 350}},\ \bibinfo {pages} {74 } (\bibinfo {year} {2013})},\
  \bibinfo {note} {100 years of Mass Spectrometry}\BibitemShut {NoStop}%
\bibitem [{\citenamefont {Kellerbauer}\ \emph {et~al.}(2003)\citenamefont
  {Kellerbauer}, \citenamefont {Blaum}, \citenamefont {Bollen}, \citenamefont
  {Herfurth}, \citenamefont {Kluge}, \citenamefont {Kuckein}, \citenamefont
  {Sauvan}, \citenamefont {Scheidenberger},\ and\ \citenamefont
  {Schweikhard}}]{kellerbauer03:_from}%
  \BibitemOpen
  \bibfield  {author} {\bibinfo {author} {\bibfnamefont {A.}~\bibnamefont
  {Kellerbauer}}, \bibinfo {author} {\bibfnamefont {K.}~\bibnamefont {Blaum}},
  \bibinfo {author} {\bibfnamefont {G.}~\bibnamefont {Bollen}}, \bibinfo
  {author} {\bibfnamefont {F.}~\bibnamefont {Herfurth}}, \bibinfo {author}
  {\bibfnamefont {H.-J.}\ \bibnamefont {Kluge}}, \bibinfo {author}
  {\bibfnamefont {M.}~\bibnamefont {Kuckein}}, \bibinfo {author} {\bibfnamefont
  {E.}~\bibnamefont {Sauvan}}, \bibinfo {author} {\bibfnamefont
  {C.}~\bibnamefont {Scheidenberger}}, \ and\ \bibinfo {author} {\bibfnamefont
  {L.}~\bibnamefont {Schweikhard}},\ }\href {\doibase
  10.1140/epjd/e2002-00222-0} {\bibfield  {journal} {\bibinfo  {journal} {The
  European Physical Journal D - Atomic, Molecular, Optical and Plasma Physics}\
  }\textbf {\bibinfo {volume} {22}},\ \bibinfo {pages} {53} (\bibinfo {year}
  {2003})}\BibitemShut {NoStop}%
\bibitem [{\citenamefont {Blaum}\ \emph {et~al.}(2004)\citenamefont {Blaum},
  \citenamefont {Beck}, \citenamefont {Bollen}, \citenamefont {Delahaye},
  \citenamefont {Gu{'e}naut}, \citenamefont {Herfurth}, \citenamefont
  {Kellerbauer}, \citenamefont {Kluge}, \citenamefont {Lunney}, \citenamefont
  {Schwarz}, \citenamefont {Schweikhard},\ and\ \citenamefont
  {Yazidjian}}]{blaum04:_popul_pennin}%
  \BibitemOpen
  \bibfield  {author} {\bibinfo {author} {\bibfnamefont {K.}~\bibnamefont
  {Blaum}}, \bibinfo {author} {\bibfnamefont {D.}~\bibnamefont {Beck}},
  \bibinfo {author} {\bibfnamefont {G.}~\bibnamefont {Bollen}}, \bibinfo
  {author} {\bibfnamefont {P.}~\bibnamefont {Delahaye}}, \bibinfo {author}
  {\bibfnamefont {C.}~\bibnamefont {Gu{'e}naut}}, \bibinfo {author}
  {\bibfnamefont {F.}~\bibnamefont {Herfurth}}, \bibinfo {author}
  {\bibfnamefont {A.}~\bibnamefont {Kellerbauer}}, \bibinfo {author}
  {\bibfnamefont {H.-J.}\ \bibnamefont {Kluge}}, \bibinfo {author}
  {\bibfnamefont {D.}~\bibnamefont {Lunney}}, \bibinfo {author} {\bibfnamefont
  {S.}~\bibnamefont {Schwarz}}, \bibinfo {author} {\bibfnamefont
  {L.}~\bibnamefont {Schweikhard}}, \ and\ \bibinfo {author} {\bibfnamefont
  {C.}~\bibnamefont {Yazidjian}},\ }\href
  {http://stacks.iop.org/0295-5075/67/i=4/a=586} {\bibfield  {journal}
  {\bibinfo  {journal} {EPL (Europhysics Letters)}\ }\textbf {\bibinfo {volume}
  {67}},\ \bibinfo {pages} {586} (\bibinfo {year} {2004})}\BibitemShut
  {NoStop}%
\bibitem [{\citenamefont {Rodrigues}\ \emph {et~al.}(2004)\citenamefont
  {Rodrigues}, \citenamefont {Indelicato}, \citenamefont {Santos},
  \citenamefont {Patt{\'e}},\ and\ \citenamefont
  {Parente}}]{rodrigues04:_system}%
  \BibitemOpen
  \bibfield  {author} {\bibinfo {author} {\bibfnamefont {G.}~\bibnamefont
  {Rodrigues}}, \bibinfo {author} {\bibfnamefont {P.}~\bibnamefont
  {Indelicato}}, \bibinfo {author} {\bibfnamefont {J.}~\bibnamefont {Santos}},
  \bibinfo {author} {\bibfnamefont {P.}~\bibnamefont {Patt{\'e}}}, \ and\
  \bibinfo {author} {\bibfnamefont {F.}~\bibnamefont {Parente}},\ }\href
  {\doibase 10.1016/j.adt.2003.11.005} {\bibfield  {journal} {\bibinfo
  {journal} {Atomic Data and Nuclear Data Tables}\ }\textbf {\bibinfo {volume}
  {86}},\ \bibinfo {pages} {117 } (\bibinfo {year} {2004})}\BibitemShut
  {NoStop}%
\bibitem [{\citenamefont {Brodeur}\ \emph {et~al.}(2009)\citenamefont
  {Brodeur}, \citenamefont {Brunner}, \citenamefont {Champagne}, \citenamefont
  {Ettenauer}, \citenamefont {Smith}, \citenamefont {Lapierre}, \citenamefont
  {Ringle}, \citenamefont {Ryjkov}, \citenamefont {Audi}, \citenamefont
  {Delheij}, \citenamefont {Lunney},\ and\ \citenamefont
  {Dilling}}]{brodeur09:_new_pennin_titan}%
  \BibitemOpen
  \bibfield  {author} {\bibinfo {author} {\bibfnamefont {M.}~\bibnamefont
  {Brodeur}}, \bibinfo {author} {\bibfnamefont {T.}~\bibnamefont {Brunner}},
  \bibinfo {author} {\bibfnamefont {C.}~\bibnamefont {Champagne}}, \bibinfo
  {author} {\bibfnamefont {S.}~\bibnamefont {Ettenauer}}, \bibinfo {author}
  {\bibfnamefont {M.}~\bibnamefont {Smith}}, \bibinfo {author} {\bibfnamefont
  {A.}~\bibnamefont {Lapierre}}, \bibinfo {author} {\bibfnamefont
  {R.}~\bibnamefont {Ringle}}, \bibinfo {author} {\bibfnamefont {V.~L.}\
  \bibnamefont {Ryjkov}}, \bibinfo {author} {\bibfnamefont {G.}~\bibnamefont
  {Audi}}, \bibinfo {author} {\bibfnamefont {P.}~\bibnamefont {Delheij}},
  \bibinfo {author} {\bibfnamefont {D.}~\bibnamefont {Lunney}}, \ and\ \bibinfo
  {author} {\bibfnamefont {J.}~\bibnamefont {Dilling}},\ }\href {\doibase
  10.1103/PhysRevC.80.044318} {\bibfield  {journal} {\bibinfo  {journal} {Phys.
  Rev. C}\ }\textbf {\bibinfo {volume} {80}},\ \bibinfo {pages} {044318}
  (\bibinfo {year} {2009})}\BibitemShut {NoStop}%
\bibitem [{\citenamefont {Wang}\ \emph {et~al.}(2012)\citenamefont {Wang},
  \citenamefont {Audi}, \citenamefont {Wapstra}, \citenamefont {Kondev},
  \citenamefont {MacCormick}, \citenamefont {Xu},\ and\ \citenamefont
  {Pfeiffer}}]{wang12:_ame20}%
  \BibitemOpen
  \bibfield  {author} {\bibinfo {author} {\bibfnamefont {M.}~\bibnamefont
  {Wang}}, \bibinfo {author} {\bibfnamefont {G.}~\bibnamefont {Audi}}, \bibinfo
  {author} {\bibfnamefont {A.}~\bibnamefont {Wapstra}}, \bibinfo {author}
  {\bibfnamefont {F.}~\bibnamefont {Kondev}}, \bibinfo {author} {\bibfnamefont
  {M.}~\bibnamefont {MacCormick}}, \bibinfo {author} {\bibfnamefont
  {X.}~\bibnamefont {Xu}}, \ and\ \bibinfo {author} {\bibfnamefont
  {B.}~\bibnamefont {Pfeiffer}},\ }\href
  {http://stacks.iop.org/1674-1137/36/i=12/a=003} {\bibfield  {journal}
  {\bibinfo  {journal} {Chinese Physics C}\ }\textbf {\bibinfo {volume} {36}},\
  \bibinfo {pages} {1603} (\bibinfo {year} {2012})}\BibitemShut {NoStop}%
\bibitem [{\citenamefont {Manea}\ \emph {et~al.}(2013)\citenamefont {Manea},
  \citenamefont {Atanasov}, \citenamefont {Beck}, \citenamefont {Blaum},
  \citenamefont {Borgmann}, \citenamefont {Cakirli}, \citenamefont {Eronen},
  \citenamefont {George}, \citenamefont {Herfurth}, \citenamefont {Herlert},
  \citenamefont {Kowalska}, \citenamefont {Kreim}, \citenamefont {Litvinov},
  \citenamefont {Lunney}, \citenamefont {Neidherr}, \citenamefont {Rosenbusch},
  \citenamefont {Schweikhard}, \citenamefont {Wienholtz}, \citenamefont
  {Wolf},\ and\ \citenamefont {Zuber}}]{manea13:_collec_rb}%
  \BibitemOpen
  \bibfield  {author} {\bibinfo {author} {\bibfnamefont {V.}~\bibnamefont
  {Manea}}, \bibinfo {author} {\bibfnamefont {D.}~\bibnamefont {Atanasov}},
  \bibinfo {author} {\bibfnamefont {D.}~\bibnamefont {Beck}}, \bibinfo {author}
  {\bibfnamefont {K.}~\bibnamefont {Blaum}}, \bibinfo {author} {\bibfnamefont
  {C.}~\bibnamefont {Borgmann}}, \bibinfo {author} {\bibfnamefont {R.~B.}\
  \bibnamefont {Cakirli}}, \bibinfo {author} {\bibfnamefont {T.}~\bibnamefont
  {Eronen}}, \bibinfo {author} {\bibfnamefont {S.}~\bibnamefont {George}},
  \bibinfo {author} {\bibfnamefont {F.}~\bibnamefont {Herfurth}}, \bibinfo
  {author} {\bibfnamefont {A.}~\bibnamefont {Herlert}}, \bibinfo {author}
  {\bibfnamefont {M.}~\bibnamefont {Kowalska}}, \bibinfo {author}
  {\bibfnamefont {S.}~\bibnamefont {Kreim}}, \bibinfo {author} {\bibfnamefont
  {Y.~A.}\ \bibnamefont {Litvinov}}, \bibinfo {author} {\bibfnamefont
  {D.}~\bibnamefont {Lunney}}, \bibinfo {author} {\bibfnamefont
  {D.}~\bibnamefont {Neidherr}}, \bibinfo {author} {\bibfnamefont
  {M.}~\bibnamefont {Rosenbusch}}, \bibinfo {author} {\bibfnamefont
  {L.}~\bibnamefont {Schweikhard}}, \bibinfo {author} {\bibfnamefont
  {F.}~\bibnamefont {Wienholtz}}, \bibinfo {author} {\bibfnamefont {R.~N.}\
  \bibnamefont {Wolf}}, \ and\ \bibinfo {author} {\bibfnamefont
  {K.}~\bibnamefont {Zuber}},\ }\href {\doibase 10.1103/PhysRevC.88.054322}
  {\bibfield  {journal} {\bibinfo  {journal} {Phys. Rev. C}\ }\textbf {\bibinfo
  {volume} {88}},\ \bibinfo {pages} {054322} (\bibinfo {year}
  {2013})}\BibitemShut {NoStop}%
\bibitem [{\citenamefont {Hager}\ \emph {et~al.}(2006)\citenamefont {Hager},
  \citenamefont {Eronen}, \citenamefont {Hakala}, \citenamefont {Jokinen},
  \citenamefont {Kolhinen}, \citenamefont {Kopecky}, \citenamefont {Moore},
  \citenamefont {Nieminen}, \citenamefont {Oinonen}, \citenamefont
  {Rinta-Antila}, \citenamefont {Szerypo},\ and\ \citenamefont
  {\"Ayst\"o}}]{hager06:_first_precis_mass_measur_refrac_fission_fragm}%
  \BibitemOpen
  \bibfield  {author} {\bibinfo {author} {\bibfnamefont {U.}~\bibnamefont
  {Hager}}, \bibinfo {author} {\bibfnamefont {T.}~\bibnamefont {Eronen}},
  \bibinfo {author} {\bibfnamefont {J.}~\bibnamefont {Hakala}}, \bibinfo
  {author} {\bibfnamefont {A.}~\bibnamefont {Jokinen}}, \bibinfo {author}
  {\bibfnamefont {V.~S.}\ \bibnamefont {Kolhinen}}, \bibinfo {author}
  {\bibfnamefont {S.}~\bibnamefont {Kopecky}}, \bibinfo {author} {\bibfnamefont
  {I.}~\bibnamefont {Moore}}, \bibinfo {author} {\bibfnamefont
  {A.}~\bibnamefont {Nieminen}}, \bibinfo {author} {\bibfnamefont
  {M.}~\bibnamefont {Oinonen}}, \bibinfo {author} {\bibfnamefont
  {S.}~\bibnamefont {Rinta-Antila}}, \bibinfo {author} {\bibfnamefont
  {J.}~\bibnamefont {Szerypo}}, \ and\ \bibinfo {author} {\bibfnamefont
  {J.}~\bibnamefont {\"Ayst\"o}},\ }\href {\doibase
  10.1103/PhysRevLett.96.042504} {\bibfield  {journal} {\bibinfo  {journal}
  {Phys. Rev. Lett.}\ }\textbf {\bibinfo {volume} {96}},\ \bibinfo {pages}
  {042504} (\bibinfo {year} {2006})}\BibitemShut {NoStop}%
\bibitem [{\citenamefont {Lhersonneau}\ \emph {et~al.}(2002)\citenamefont
  {Lhersonneau}, \citenamefont {Pfeiffer}, \citenamefont {Capote},
  \citenamefont {Quesada}, \citenamefont {Gabelmann},\ and\ \citenamefont
  {Kratz}}]{lhersonneau02:_k}%
  \BibitemOpen
  \bibfield  {author} {\bibinfo {author} {\bibfnamefont {G.}~\bibnamefont
  {Lhersonneau}}, \bibinfo {author} {\bibfnamefont {B.}~\bibnamefont
  {Pfeiffer}}, \bibinfo {author} {\bibfnamefont {R.}~\bibnamefont {Capote}},
  \bibinfo {author} {\bibfnamefont {J.~M.}\ \bibnamefont {Quesada}}, \bibinfo
  {author} {\bibfnamefont {H.}~\bibnamefont {Gabelmann}}, \ and\ \bibinfo
  {author} {\bibfnamefont {K.-L.}\ \bibnamefont {Kratz}},\ }\href {\doibase
  10.1103/PhysRevC.65.024318} {\bibfield  {journal} {\bibinfo  {journal} {Phys.
  Rev. C}\ }\textbf {\bibinfo {volume} {65}},\ \bibinfo {pages} {024318}
  (\bibinfo {year} {2002})}\BibitemShut {NoStop}%
\bibitem [{\citenamefont {Gallant}\ \emph {et~al.}(2012)\citenamefont
  {Gallant}, \citenamefont {Brodeur}, \citenamefont {Brunner}, \citenamefont
  {Chowdhury}, \citenamefont {Ettenauer}, \citenamefont {Simon}, \citenamefont
  {Man{\'e}}, \citenamefont {Simon}, \citenamefont {Andreoiu}, \citenamefont
  {Delheij}, \citenamefont {Gwinner}, \citenamefont {Pearson}, \citenamefont
  {Ringle},\ and\ \citenamefont {Dilling}}]{gallant12:_highl_pennin}%
  \BibitemOpen
  \bibfield  {author} {\bibinfo {author} {\bibfnamefont {A.~T.}\ \bibnamefont
  {Gallant}}, \bibinfo {author} {\bibfnamefont {M.}~\bibnamefont {Brodeur}},
  \bibinfo {author} {\bibfnamefont {T.}~\bibnamefont {Brunner}}, \bibinfo
  {author} {\bibfnamefont {U.}~\bibnamefont {Chowdhury}}, \bibinfo {author}
  {\bibfnamefont {S.}~\bibnamefont {Ettenauer}}, \bibinfo {author}
  {\bibfnamefont {V.~V.}\ \bibnamefont {Simon}}, \bibinfo {author}
  {\bibfnamefont {E.}~\bibnamefont {Man{\'e}}}, \bibinfo {author}
  {\bibfnamefont {M.~C.}\ \bibnamefont {Simon}}, \bibinfo {author}
  {\bibfnamefont {C.}~\bibnamefont {Andreoiu}}, \bibinfo {author}
  {\bibfnamefont {P.}~\bibnamefont {Delheij}}, \bibinfo {author} {\bibfnamefont
  {G.}~\bibnamefont {Gwinner}}, \bibinfo {author} {\bibfnamefont {M.~R.}\
  \bibnamefont {Pearson}}, \bibinfo {author} {\bibfnamefont {R.}~\bibnamefont
  {Ringle}}, \ and\ \bibinfo {author} {\bibfnamefont {J.}~\bibnamefont
  {Dilling}},\ }\href {\doibase 10.1103/PhysRevC.85.044311} {\bibfield
  {journal} {\bibinfo  {journal} {Phys. Rev. C}\ }\textbf {\bibinfo {volume}
  {85}},\ \bibinfo {pages} {044311} (\bibinfo {year} {2012})}\BibitemShut
  {NoStop}%
\bibitem [{\citenamefont {Procter}\ \emph {et~al.}(2015)\citenamefont
  {Procter}, \citenamefont {Behr}, \citenamefont {Billowes}, \citenamefont
  {Buchinger}, \citenamefont {Cheal}, \citenamefont {Crawford}, \citenamefont
  {Dilling}, \citenamefont {Garnsworthy}, \citenamefont {Leary}, \citenamefont
  {Levy}, \citenamefont {Man{\'e}}, \citenamefont {Pearson}, \citenamefont
  {Shelbaya}, \citenamefont {Stolz}, \citenamefont {Al~Tamimi},\ and\
  \citenamefont {Voss}}]{procter15:_direc}%
  \BibitemOpen
  \bibfield  {author} {\bibinfo {author} {\bibfnamefont {T.}~\bibnamefont
  {Procter}}, \bibinfo {author} {\bibfnamefont {J.}~\bibnamefont {Behr}},
  \bibinfo {author} {\bibfnamefont {J.}~\bibnamefont {Billowes}}, \bibinfo
  {author} {\bibfnamefont {F.}~\bibnamefont {Buchinger}}, \bibinfo {author}
  {\bibfnamefont {B.}~\bibnamefont {Cheal}}, \bibinfo {author} {\bibfnamefont
  {J.}~\bibnamefont {Crawford}}, \bibinfo {author} {\bibfnamefont
  {J.}~\bibnamefont {Dilling}}, \bibinfo {author} {\bibfnamefont
  {A.}~\bibnamefont {Garnsworthy}}, \bibinfo {author} {\bibfnamefont
  {A.}~\bibnamefont {Leary}}, \bibinfo {author} {\bibfnamefont
  {C.}~\bibnamefont {Levy}}, \bibinfo {author} {\bibfnamefont {E.}~\bibnamefont
  {Man{\'e}}}, \bibinfo {author} {\bibfnamefont {M.}~\bibnamefont {Pearson}},
  \bibinfo {author} {\bibfnamefont {O.}~\bibnamefont {Shelbaya}}, \bibinfo
  {author} {\bibfnamefont {M.}~\bibnamefont {Stolz}}, \bibinfo {author}
  {\bibfnamefont {W.}~\bibnamefont {Al~Tamimi}}, \ and\ \bibinfo {author}
  {\bibfnamefont {A.}~\bibnamefont {Voss}},\ }\href {\doibase
  10.1140/epja/i2015-15023-2} {\bibfield  {journal} {\bibinfo  {journal} {The
  European Physical Journal A}\ }\textbf {\bibinfo {volume} {51}},\ \bibinfo
  {eid} {23} (\bibinfo {year} {2015}),\ 10.1140/epja/i2015-15023-2}\BibitemShut
  {NoStop}%
\bibitem [{\citenamefont {Audi}\ \emph {et~al.}(2012)\citenamefont {Audi},
  \citenamefont {Wang}, \citenamefont {Wapstra}, \citenamefont {Kondev},
  \citenamefont {MacCormick}, \citenamefont {Xu},\ and\ \citenamefont
  {Pfeiffer}}]{audi12:_ame20}%
  \BibitemOpen
  \bibfield  {author} {\bibinfo {author} {\bibfnamefont {G.}~\bibnamefont
  {Audi}}, \bibinfo {author} {\bibfnamefont {M.}~\bibnamefont {Wang}}, \bibinfo
  {author} {\bibfnamefont {A.}~\bibnamefont {Wapstra}}, \bibinfo {author}
  {\bibfnamefont {F.}~\bibnamefont {Kondev}}, \bibinfo {author} {\bibfnamefont
  {M.}~\bibnamefont {MacCormick}}, \bibinfo {author} {\bibfnamefont
  {X.}~\bibnamefont {Xu}}, \ and\ \bibinfo {author} {\bibfnamefont
  {B.}~\bibnamefont {Pfeiffer}},\ }\href
  {http://stacks.iop.org/1674-1137/36/i=12/a=002} {\bibfield  {journal}
  {\bibinfo  {journal} {Chinese Physics C}\ }\textbf {\bibinfo {volume} {36}},\
  \bibinfo {pages} {1287} (\bibinfo {year} {2012})}\BibitemShut {NoStop}%
\bibitem [{\citenamefont {Audi}\ \emph {et~al.}(1986)\citenamefont {Audi},
  \citenamefont {Coc}, \citenamefont {Epherre-Rey-Campagnolle}, \citenamefont
  {Scornet}, \citenamefont {Thibault},\ and\ \citenamefont
  {Touchard}}]{audi86:_mass_rb_cs_fr}%
  \BibitemOpen
  \bibfield  {author} {\bibinfo {author} {\bibfnamefont {G.}~\bibnamefont
  {Audi}}, \bibinfo {author} {\bibfnamefont {A.}~\bibnamefont {Coc}}, \bibinfo
  {author} {\bibfnamefont {M.}~\bibnamefont {Epherre-Rey-Campagnolle}},
  \bibinfo {author} {\bibfnamefont {G.~L.}\ \bibnamefont {Scornet}}, \bibinfo
  {author} {\bibfnamefont {C.}~\bibnamefont {Thibault}}, \ and\ \bibinfo
  {author} {\bibfnamefont {F.}~\bibnamefont {Touchard}},\ }\href {\doibase
  10.1016/0375-9474(86)90232-0} {\bibfield  {journal} {\bibinfo  {journal}
  {Nuclear Physics A}\ }\textbf {\bibinfo {volume} {449}},\ \bibinfo {pages}
  {491 } (\bibinfo {year} {1986})}\BibitemShut {NoStop}%
\bibitem [{\citenamefont {Iafigliola}\ \emph {et~al.}(1984)\citenamefont
  {Iafigliola}, \citenamefont {Dautet}, \citenamefont {Xu}, \citenamefont
  {Lee}, \citenamefont {Chrien}, \citenamefont {Gill},\ and\ \citenamefont
  {Shmid}}]{iafigliola84:_beta_rb_sr}%
  \BibitemOpen
  \bibfield  {author} {\bibinfo {author} {\bibfnamefont {R.}~\bibnamefont
  {Iafigliola}}, \bibinfo {author} {\bibfnamefont {H.}~\bibnamefont {Dautet}},
  \bibinfo {author} {\bibfnamefont {S.~W.}\ \bibnamefont {Xu}}, \bibinfo
  {author} {\bibfnamefont {J.~K.~P.}\ \bibnamefont {Lee}}, \bibinfo {author}
  {\bibfnamefont {R.}~\bibnamefont {Chrien}}, \bibinfo {author} {\bibfnamefont
  {R.}~\bibnamefont {Gill}}, \ and\ \bibinfo {author} {\bibfnamefont
  {M.}~\bibnamefont {Shmid}},\ }\href@noop {} {\bibfield  {journal} {\bibinfo
  {journal} {Proceedings of the 7th International, Conference Atomic Masses and
  Fundamental Constants AMCO--7}\ ,\ \bibinfo {pages} {141}} (\bibinfo {year}
  {1984})}\BibitemShut {NoStop}%
\bibitem [{\citenamefont {Balog}\ \emph {et~al.}(1992)\citenamefont {Balog},
  \citenamefont {Graefenstedt}, \citenamefont {Gro{\ss}}, \citenamefont
  {J{\"u}rgens}, \citenamefont {Keyser}, \citenamefont {M{\"u}nnich},
  \citenamefont {Otto}, \citenamefont {Schreiber}, \citenamefont {Winkelmann},\
  and\ \citenamefont {Wulff}}]{balog92:_exper}%
  \BibitemOpen
  \bibfield  {author} {\bibinfo {author} {\bibfnamefont {K.}~\bibnamefont
  {Balog}}, \bibinfo {author} {\bibfnamefont {M.}~\bibnamefont {Graefenstedt}},
  \bibinfo {author} {\bibfnamefont {M.}~\bibnamefont {Gro{\ss}}}, \bibinfo
  {author} {\bibfnamefont {P.}~\bibnamefont {J{\"u}rgens}}, \bibinfo {author}
  {\bibfnamefont {U.}~\bibnamefont {Keyser}}, \bibinfo {author} {\bibfnamefont
  {F.}~\bibnamefont {M{\"u}nnich}}, \bibinfo {author} {\bibfnamefont
  {T.}~\bibnamefont {Otto}}, \bibinfo {author} {\bibfnamefont {F.}~\bibnamefont
  {Schreiber}}, \bibinfo {author} {\bibfnamefont {T.}~\bibnamefont
  {Winkelmann}}, \ and\ \bibinfo {author} {\bibfnamefont {J.}~\bibnamefont
  {Wulff}},\ }\href {\doibase 10.1007/BF01288459} {\bibfield  {journal}
  {\bibinfo  {journal} {Zeitschrift f{\"u}r Physik A Hadrons and Nuclei}\
  }\textbf {\bibinfo {volume} {342}},\ \bibinfo {pages} {125} (\bibinfo {year}
  {1992})}\BibitemShut {NoStop}%
\bibitem [{\citenamefont {Van~Schelt}\ \emph {et~al.}(2012)\citenamefont
  {Van~Schelt}, \citenamefont {Lascar}, \citenamefont {Savard}, \citenamefont
  {Clark}, \citenamefont {Caldwell}, \citenamefont {Chaudhuri}, \citenamefont
  {Fallis}, \citenamefont {Greene}, \citenamefont {Levand}, \citenamefont {Li},
  \citenamefont {Sharma}, \citenamefont {Sternberg}, \citenamefont {Sun},\ and\
  \citenamefont {Zabransky}}]{van12:_mass_canad_pennin_trap}%
  \BibitemOpen
  \bibfield  {author} {\bibinfo {author} {\bibfnamefont {J.}~\bibnamefont
  {Van~Schelt}}, \bibinfo {author} {\bibfnamefont {D.}~\bibnamefont {Lascar}},
  \bibinfo {author} {\bibfnamefont {G.}~\bibnamefont {Savard}}, \bibinfo
  {author} {\bibfnamefont {J.~A.}\ \bibnamefont {Clark}}, \bibinfo {author}
  {\bibfnamefont {S.}~\bibnamefont {Caldwell}}, \bibinfo {author}
  {\bibfnamefont {A.}~\bibnamefont {Chaudhuri}}, \bibinfo {author}
  {\bibfnamefont {J.}~\bibnamefont {Fallis}}, \bibinfo {author} {\bibfnamefont
  {J.~P.}\ \bibnamefont {Greene}}, \bibinfo {author} {\bibfnamefont {A.~F.}\
  \bibnamefont {Levand}}, \bibinfo {author} {\bibfnamefont {G.}~\bibnamefont
  {Li}}, \bibinfo {author} {\bibfnamefont {K.~S.}\ \bibnamefont {Sharma}},
  \bibinfo {author} {\bibfnamefont {M.~G.}\ \bibnamefont {Sternberg}}, \bibinfo
  {author} {\bibfnamefont {T.}~\bibnamefont {Sun}}, \ and\ \bibinfo {author}
  {\bibfnamefont {B.~J.}\ \bibnamefont {Zabransky}},\ }\href {\doibase
  10.1103/PhysRevC.85.045805} {\bibfield  {journal} {\bibinfo  {journal} {Phys.
  Rev. C}\ }\textbf {\bibinfo {volume} {85}},\ \bibinfo {pages} {045805}
  (\bibinfo {year} {2012})}\BibitemShut {NoStop}%
\bibitem [{\citenamefont {Rauscher}\ and\ \citenamefont
  {Thielemann}(2000)}]{rauscher00:_astrop_react_rates_from_statis_model_calcul}%
  \BibitemOpen
  \bibfield  {author} {\bibinfo {author} {\bibfnamefont {T.}~\bibnamefont
  {Rauscher}}\ and\ \bibinfo {author} {\bibfnamefont {F.-K.}\ \bibnamefont
  {Thielemann}},\ }\href {\doibase 10.1006/adnd.2000.0834} {\bibfield
  {journal} {\bibinfo  {journal} {Atomic Data and Nuclear Data Tables}\
  }\textbf {\bibinfo {volume} {75}},\ \bibinfo {pages} {1 } (\bibinfo {year}
  {2000})}\BibitemShut {NoStop}%
\bibitem [{\citenamefont {Ke}\ \emph {et~al.}(2006)\citenamefont {Ke},
  \citenamefont {Shi}, \citenamefont {Gwinner}, \citenamefont {Sharma},
  \citenamefont {Toews}, \citenamefont {Dilling},\ and\ \citenamefont
  {Ryjkov}}]{ke06:_titan_trium}%
  \BibitemOpen
  \bibfield  {author} {\bibinfo {author} {\bibfnamefont {Z.}~\bibnamefont
  {Ke}}, \bibinfo {author} {\bibfnamefont {W.}~\bibnamefont {Shi}}, \bibinfo
  {author} {\bibfnamefont {G.}~\bibnamefont {Gwinner}}, \bibinfo {author}
  {\bibfnamefont {K.}~\bibnamefont {Sharma}}, \bibinfo {author} {\bibfnamefont
  {S.}~\bibnamefont {Toews}}, \bibinfo {author} {\bibfnamefont
  {J.}~\bibnamefont {Dilling}}, \ and\ \bibinfo {author} {\bibfnamefont
  {V.}~\bibnamefont {Ryjkov}},\ }\href {\doibase 10.1007/s10751-007-9548-x}
  {\bibfield  {journal} {\bibinfo  {journal} {Hyperfine Interactions}\ }\textbf
  {\bibinfo {volume} {173}},\ \bibinfo {pages} {103} (\bibinfo {year}
  {2006})}\BibitemShut {NoStop}%
\bibitem [{\citenamefont {Jesch}\ \emph {et~al.}(2015)\citenamefont {Jesch},
  \citenamefont {Dickel}, \citenamefont {Pla{\ss}}, \citenamefont {Short},
  \citenamefont {Ayet San~Andres}, \citenamefont {Dilling}, \citenamefont
  {Geissel}, \citenamefont {Greiner}, \citenamefont {Lang}, \citenamefont
  {Leach}, \citenamefont {Lippert}, \citenamefont {Scheidenberger},\ and\
  \citenamefont {Yavor}}]{jesch15:_mr_tof_ms_titan_trium}%
  \BibitemOpen
  \bibfield  {author} {\bibinfo {author} {\bibfnamefont {C.}~\bibnamefont
  {Jesch}}, \bibinfo {author} {\bibfnamefont {T.}~\bibnamefont {Dickel}},
  \bibinfo {author} {\bibfnamefont {W.~R.}\ \bibnamefont {Pla{\ss}}}, \bibinfo
  {author} {\bibfnamefont {D.}~\bibnamefont {Short}}, \bibinfo {author}
  {\bibfnamefont {S.}~\bibnamefont {Ayet San~Andres}}, \bibinfo {author}
  {\bibfnamefont {J.}~\bibnamefont {Dilling}}, \bibinfo {author} {\bibfnamefont
  {H.}~\bibnamefont {Geissel}}, \bibinfo {author} {\bibfnamefont
  {F.}~\bibnamefont {Greiner}}, \bibinfo {author} {\bibfnamefont
  {J.}~\bibnamefont {Lang}}, \bibinfo {author} {\bibfnamefont {K.~G.}\
  \bibnamefont {Leach}}, \bibinfo {author} {\bibfnamefont {W.}~\bibnamefont
  {Lippert}}, \bibinfo {author} {\bibfnamefont {C.}~\bibnamefont
  {Scheidenberger}}, \ and\ \bibinfo {author} {\bibfnamefont {M.~I.}\
  \bibnamefont {Yavor}},\ }\href {\doibase 10.1007/s10751-015-1184-2}
  {\bibfield  {journal} {\bibinfo  {journal} {Hyperfine Interactions}\ ,\
  \bibinfo {pages} {1}} (\bibinfo {year} {2015})}\BibitemShut {NoStop}%
\end{thebibliography}%
\end{document}